\documentclass[11pt, a4paper]{article}
\usepackage{jheppub}
\usepackage{bbm}
\usepackage{amsmath}
\usepackage{amssymb}
\usepackage{amsfonts}
\usepackage{mathrsfs}
\usepackage{graphicx}
\usepackage[11pt]{moresize}
\usepackage[normalem]{ulem}
\usepackage[dvipsnames]{xcolor}
\usepackage[]{hyperref}
\usepackage{orcidlink}
\usepackage{relsize}
\usepackage{adjustbox}
\usepackage[verbose]{placeins}
\usepackage{multirow}
\usepackage{float}
\usepackage{footnote}
\usepackage{tablefootnote}
\usepackage{lineno}
\usepackage{makecell}
\hypersetup{
 bookmarks  = true,      
 unicode    = false,     
 pdfcreator = {RevTeX},  
 colorlinks = true,      
 linkcolor  = red,       
 citecolor  = blue,      
 filecolor  = black,     
 urlcolor   = blue,      
}

\newcommand{\capdef}{}
\newcommand{\mycaption}[2][\capdef]{\renewcommand{\capdef}{#2}
\caption[#1]{{\footnotesize #2}}}
\makeatletter
\renewcommand{\fnum@table}{\textbf{\tablename~\thetable}}
\renewcommand{\fnum@figure}{\textbf{\figurename~\thefigure}}

\newcommand{\ldm}{\ensuremath{\Delta m_{31}^2}}

\newcommand{\tym}{\ensuremath{\theta_{13}}}
\newcommand{\tzm}{\ensuremath{\theta_{23}}}
\newcommand{\aem}{\ensuremath{a_{e\mu}}}

\newcommand{\aet}{\ensuremath{a_{e\tau}}}
\newcommand{\amt}{\ensuremath{a_{\mu\tau}}}

\newcommand{\cem}{\ensuremath{c_{e\mu}}}

\newcommand{\cet}{\ensuremath{c_{e\tau}}}
\newcommand{\cmt}{\ensuremath{c_{\mu\tau}}}

\newcommand{\ie}{\textit{i.e.}}

\makeatother

\newenvironment{conditions*}
{\par\vspace{\abovedisplayskip}\noindent
  \tabularx{\columnwidth}{>{$}l<{$} @{${}={}$} >{\raggedright\arraybackslash}X}}
{\endtabularx\par\vspace{\belowdisplayskip}}

\preprint{IP/BBSR/2023-01}

\title{Constraining Lorentz Invariance Violation with Next-Generation Long-Baseline Experiments}

\author[a, b, c]{\orcidlink{0000-0002-9714-8866}Sanjib Kumar Agarwalla,}
\author[a, b]   {\orcidlink{0000-0002-5508-7751}Sudipta Das,}
\author[a, b]   {\orcidlink{0000-0001-6719-7723}Sadashiv Sahoo,}
\author[a, b]   {\orcidlink{0000-0003-3008-480X}Pragyanprasu Swain}

\affiliation[a]{Institute of Physics, Sachivalaya Marg, Sainik School Post, Bhubaneswar 751005, India}
\affiliation[b]{Homi Bhabha National Institute, Anushakti Nagar, Mumbai 400094, India}
\affiliation[c]{Department of Physics \& Wisconsin IceCube Particle Astrophysics Center,\\ University of Wisconsin, Madison, WI 53706, U.S.A}

\emailAdd{sanjib@iopb.res.in}
\emailAdd{sudipta.d@iopb.res.in}
\emailAdd{sadashiv.sahoo@iopb.res.in}
\emailAdd{pragyanprasu.s@iopb.res.in}

\abstract
{Unified theories such as string theory and loop quantum gravity allow the Lorentz Invariance Violation (LIV) at the Planck Scale ($M_P \sim 10^{19}$ GeV). Using an effective field theory, this effect can be observed at low energies in terms of new interactions with a strength of $\sim 1/M_P$. These new interactions contain operators with LIV coefficients which can be CPT-violating or CPT-conserving. In this work, we study in detail how these LIV parameters modify the transition probabilities in the next-generation long-baseline experiments, DUNE and Hyper-K. We evaluate the sensitivities of these experiments in isolation and combination to constrain the off-diagonal CPT-violating ($a_{e\mu}$, $a_{e\tau}$, $a_{\mu\tau}$) and CPT-conserving ($c_{e\mu}$, $c_{e\tau}$, $c_{\mu\tau}$) LIV parameters. We derive approximate compact analytical expressions of appearance ($\nu_{\mu}\to\nu_e$) and disappearance ($\nu_{\mu}\to\nu_\mu$) probabilities in the presence of these LIV parameters to explain our numerical results. We explore the possible correlations and degeneracies between these LIV parameters and the most uncertain 3$\nu$ oscillation parameters, namely, $\theta_{23}$ and $\delta_{\rm CP}$. We find that for non-maximal values of $\theta_{23}$ ($\theta_{23}\neq45^\circ$), there exist degenerate solutions in its opposite octant for standalone DUNE and Hyper-K. These degeneracies disappear when we combine the data from DUNE and Hyper-K. In case of no-show, we place the expected upper bounds on these CPT-violating and CPT-conserving LIV parameters at 95\% C.L. using the standalone DUNE, Hyper-K, and their combination. We observe that due to its access to a longer baseline and multi-GeV neutrinos, 
DUNE has a better reach in probing all these LIV parameters as compared to Hyper-K. Since the terms containing the CPT-conserving LIV parameters are proportional to neutrino energy in oscillation probabilities, Hyper-K is almost insensitive to the CPT-conserving LIV parameters because it mostly deals with sub-GeV neutrinos.} 

\arxivnumber{2302.12005}

\begin{document}
\maketitle
\flushbottom

\section{Introduction and Motivation}
\label{sec:introduction}

The Standard Model (SM) of particle physics has been the most successful theory of elementary particles that provides excellent explanations of many physical phenomena occurring in Nature~\cite{Workman:2022ynf}. However, there are a plethora of motivations for physicists to go beyond the SM, for example, to explain the observed matter-antimatter asymmetry in the Universe, the existence of dark matter and dark energy, and the non-zero neutrino mass. The observed mass-induced flavor transition~\cite{Super-Kamiokande:2004orf, Mohapatra:2005wg, Strumia:2006db, Gonzalez-Garcia:2007dlo, Fantini:2018itu} that requires the neutrinos to be massive provides the first experimental signature of physics beyond the SM. The standard three-flavor neutrino oscillation framework involves three mixing angles ($\theta_{12}$, $\theta_{13}$, and $\theta_{23}$), one Dirac CP phase ($\delta_{\rm CP}$), and two independent mass-squared differences, $\Delta{m}^2_{21}~ (\equiv{m}^2_2-{m}^2_1~\text{in the solar sector})$ and $\Delta{m}^2_{31}~(\equiv{m}^3_3-{m}^2_1~\text{in the atmosperic sector})$. Now that the phenomenon of neutrino oscillation has been well established, the focus has been shifted to measure the oscillation parameters with utmost precision. Marvelous data from various neutrino oscillation experiments such as Super-K-Solar~\cite{Super-Kamiokande:2016yck}, SNO~\cite{SNO:2011hxd}, BOREXINO~\cite{BOREXINO:2014pcl}, Super-K-Atmospheric~\cite{Super-Kamiokande:2004orf, Super-Kamiokande:2010tar, Super-Kamiokande:2017yvm, Super-Kamiokande:2019gzr}, IceCube-DeepCore~\cite{IceCube:2017lak}, ANTARES~\cite{ANTARES:2018rtf}, Daya Bay~\cite{DayaBay:2018yms}, RENO~\cite{RENO:2018dro}, Double Chooz~\cite{DoubleChooz:2019qbj}, MINOS~\cite{MINOS:2013utc}, Tokai-to-Kamioka (T2K)~\cite{T2K:2019bcf}, and NuMI Off-axis $\nu_{e}$ Appearance (NO$\nu$A)~\cite{NOvA:2019cyt, NOvA:2021nfi} have already provided a first order picture of the lepton mixing pattern in three-flavor scenario.

There are three major issues in the three-flavor neutrino oscillation paradigm that are yet to be resolved, namely, the value of the CP phase ($\delta_{\rm CP}$), the octant of the atmospheric mixing angle ($\theta_{23}$), and the neutrino mass ordering. The future long-baseline (LBL) experiment, Deep Underground Neutrino Experiment (DUNE)~\cite{DUNE:2015lol, DUNE:2020lwj, DUNE:2020ypp, DUNE:2020jqi, DUNE:2021cuw, DUNE:2021mtg} with its wide band neutrino beam, will play a crucial role in establishing the deviation of the atmospheric mixing angle ($\theta_{23}$) from its maximal value and settling down its correct octant with outstanding precision~\cite{Agarwalla:2021bzs}. DUNE can measure the value of atmospheric mass splitting at several $L/E$ values and settle the issue of neutrino mass ordering at a high confidence level exploiting the Earth's matter effect that it possesses due to its long baseline~\cite{DUNE:2020ypp}. DUNE is also capable to establish leptonic CP violation by measuring the value of $\delta_{\rm CP}$ precisely~\cite{Agarwalla:2022xdo}. Another proposed long-baseline experiment which spans from Tokai to Hyper-Kamiokande (abbreviated as Hyper-K from now onwards)~\cite{Hyper-KamiokandeWorkingGroup:2014czz, Abe:2015zbg, Hyper-Kamiokande:2018ofw} will also shed light on these pressing issues. In the Hyper-K setup, the detector is placed in Japan, which is 295 km away from the J-PARC facility and will receive a highly intense narrow-band neutrino beam from the J-PARC source at an off-axis angle of $2.5^\circ$. Hyper-K will have a four times smaller baseline than DUNE, which in turn will face negligible Earth's matter effect and hence can measure the value of $\delta_{\rm CP}$ and establish leptonic CP violation without having the interference of the fake CP-asymmetry induced by Earth's matter~\cite{Agarwalla:2022xdo}.

Apart from measuring the standard oscillation parameters with high precision, the long-baseline experiments will also search for new physics beyond the Standard Model (BSM)\footnote{For an extensive review on this topic, see Refs.~\cite{Arguelles:2019xgp, Arguelles:2022tki}.} through neutrino oscillation, namely, eV-scale sterile neutrino~\cite{Berryman:2015nua, Agarwalla:2016mrc, Agarwalla:2016xxa, Agarwalla:2016xlg, Agarwalla:2018nlx, KumarAgarwalla:2019blx}, neutrino non-standard interactions~\cite{Coloma:2015kiu, Agarwalla:2016fkh}, non-unitary neutrino mixing~\cite{Escrihuela:2016ube, Agarwalla:2021owd}, long-range interactions~\cite{Chatterjee:2015gta}, neutrino decay~\cite{Choubey:2017dyu, Coloma:2017zpg}, and Lorentz Invariance Violation (LIV)~\cite{Barenboim:2018ctx, KumarAgarwalla:2019gdj, Fiza:2022xfw}. In this work, we mainly focus on LIV. It is a well-established fact that the Lorentz symmetry is an exact symmetry of Nature. As a consequence, the Standard Model of particle physics conserves Lorentz symmetry. However, there exist several models, unifying SM and general relativity, that violate the Lorentz and CPT symmetry at the Planck scale ($\sim10^{19}$ GeV). This can be realized at a low energy scale accessible to the current experiments under Standard Model Extension (SME) framework. Various neutrino experiments are at the forefront to test the Lorentz Invariance Violation at a low energy scale. For example, in an attempt to understand the excess of $\nu_e$ signal events in the $\nu_{\mu}$ beam, LSND collaboration~\cite{LSND:2005oop}, searched for the possible signature of LIV in the context of neutrino oscillation. They did not find any signature of LIV and put strong constraints on relevant LIV parameters.

Several other experiments have made efforts to search for LIV, which include MINOS~\cite{MINOS:2008fnv, MINOS:2010kat, MINOS:2012ozn}, MiniBooNE~\cite{MiniBooNE:2011pix}, Double Chooz~\cite{DoubleChooz:2012eiq}, Super-K~\cite{Super-Kamiokande:2014exs}, IceCube~\cite{IceCube:2010fyu} and T2K~\cite{Abe:2017eot}. None of these experiments found any positive signal of LIV and set competitive bounds on various LIV parameters. In addition to the aforementioned studies by the experimental collaborations, there are various independent works towards the exploration of LIV with accelerator neutrinos in long-baseline experiments~\cite{Dighe:2008bu, Barenboim:2009ts, Rebel:2013vc, Diaz:2015dxa, deGouvea:2017yvn, Barenboim:2017ewj, Barenboim:2018lpo, Barenboim:2018ctx, Majhi:2019tfi, Fiza:2022xfw, Majhi:2022fed}, reactor antineutrinos in short-baseline experiments~\cite{Giunti:2010zs}, atmospheric neutrinos~\cite{Datta:2003dg, Chatterjee:2014oda, SinghKoranga:2014mxh, Sahoo:2021dit, Sahoo:2022rns}, solar neutrinos~\cite{Diaz:2016fqd}, and high-energy neutrinos from astrophysical sources~\cite{Hooper:2005jp, Tomar:2015fha, Liao:2017yuy}. Recently, the KATRIN experiment, using the data from the first scientific run, placed limits on some of the oscillation-free LIV parameters that can't be probed by the time-of-flight or neutrino oscillation experiments~\cite{KATRIN:2022qou}. Note that due to $SU(2)_{L}$ gauge invariance in the electroweak sector of the SME Lagrangian, the bounds on the LIV parameters obtained from the neutrino experiments and the charged-lepton sector may be related to each other~\cite{Crivellin:2020oov}. An exhaustive list of the constraints on all the CPT-violating and CPT-conserving LIV parameters can be found in Ref.~\cite{Kostelecky:2008ts}.

In the present work, we mainly focus on the capability of the upcoming long-baseline experiments, DUNE and Hyper-K in isolation and combination, to constrain the LIV parameters. We derive the sensitivities of these experiments to place competitive limits on the off-diagonal CPT-violating LIV parameters ($a_{\alpha\beta}$ where $\alpha, \beta = e, \mu, \tau$ and $\alpha\neq\beta$) and for the first time, the off-diagonal CPT-conserving LIV parameters ($c_{\alpha\beta}$ where $\alpha, \beta = e, \mu, \tau$ and $\alpha\neq\beta$). We study the impact of these LIV parameters and their associated phases at the probability level. Then we shift our attention to explore the possible degeneracies between the standard oscillation parameters ($\theta_{23}$ and $\delta_{\rm CP}$) and the above-mentioned LIV parameters. Finally, we derive the expected constraints on these LIV parameters using the upcoming long-baseline experiments, DUNE and Hyper-K in standalone mode and also in combination, considering their state-of-the-art simulation details. To understand various interesting features of our numerical results, we derive simple and compact analytical expressions of the oscillation probabilities for both appearance and disappearance channels.

This paper is organized as follows. In section~\ref{sec:LIV}, we discuss the formalism of neutrino oscillation in the presence of Lorentz Invariance Violation and the effects of various LIV parameters on the appearance and disappearance probabilities. In section~\ref{sec:LBL}, we give the details of the long-baseline experimental setups considered for our work and discuss the expected synergies between DUNE and Hyper-K in various aspects. Also, this section discusses the effect of LIV parameters at the event level. We dedicate section~\ref{sec:RnA} to describe the numerical technique used for our analyses. We present our results in section~\ref{sec:results}, where we show the correlations among different LIV parameters, the atmospheric mixing angle ($\theta_{23}$), and the Dirac CP phase ($\delta_{\rm CP}$) and finally the expected bounds on the CPT-conserving and CPT-violating LIV parameters. We summarize our results and give our concluding remarks in section~\ref{sec:SnC}. In appendix~\ref{appndx}, we compare the numerical (exact) and analytical (approximate) probabilities. In appendix~\ref{appndx-B}, we discuss the neutrino appearance and disappearance event spectra for DUNE and Hyper-K in standard interaction case and with new physics.
\section{Neutrino Oscillation in Presence of Lorentz Invariance Violation}
\label{sec:LIV}
\subsection{Theoretical Formalism of LIV}
The Lorentz invariance has been considered to be an inalienable part in both the Standard Model of particle physics and General Relativity for the global as well as the local variables. However, a few proposed models in String Theory~\cite{Polyakov:1987ez, Kostelecky:1988zi, Kostelecky:1989jp, Kostelecky:1990pe, Kostelecky:1991ak, Kostelecky:1995qk, Kostelecky:1999mu, Kostelecky:2000hz} and Loop Quantum Gravity~\cite{Gambini:1998it, Alfaro:2002xz, Sudarsky:2002ue, Amelino-Camelia:2002aqz, Ng:2003jk} allow the Lorentz invariance violation while attempting a unification of gravity with the SM gauge fields at the Planck scale ($M_P \sim 10^{19}$ GeV). Here, we consider the mechanism proposed in the string theory, which spontaneously breaks the CPT and Lorentz symmetries at a higher dimension $(>4)$ of space-time. One can realize such a violation of Lorentz and CPT symmetries in the realistic four-dimensional space-time by introducing new interactions in the minimal SME framework in the form of tiny perturbations. In this minimal SME framework, the strength of such interactions are expected to be suppressed by  $\sim1/M_P$~\cite{Colladay:1998fq, Kostelecky:2003fs, Colladay:1996iz, Kostelecky:2000mm, Kostelecky:2003cr, Bluhm:2005uj}, manifesting the effect of Planck scale physics at low energy. The impacts of such LIV interactions can be experienced by the fundamental particles in a broad category of experiments via coherent, interference, or extreme effects.

By virtue of mass-induced neutrino flavor oscillations, the neutrinos are sensitive to the LIV effects while propagating through space-time. Under the minimal SM extension (SME), the Lagrangian density of the induced renormalizable and gauge-invariant LIV interaction terms for the left-handed neutrinos can be expressed as~\cite{Kostelecky:2011gq, KumarAgarwalla:2019gdj, Antonelli:2020nhn, Sahoo:2021dit, Sahoo:2022rns}:
\begin{align}
	\mathcal{L}_{\rm LIV} & = -\frac{1}{2}\left[a^{\mu}_{\alpha\beta}\,\overline{\psi}_\alpha\,\gamma_{\mu}\,P_L\,\psi_\beta - i c^{\mu\nu}_{\alpha\beta}\,\overline{\psi}_\alpha\,\gamma_{\mu}\,\partial_\nu P_L\,\psi_\beta \right] + h.c.\,, 
	\label{Eq:LIV-1}
\end{align}
where $P_L$ is the projection operator, $a^{\mu}$ and $c^{\mu\nu}$ are the CPT-violating and CPT-conserving LIV parameters, respectively. Here, $(\mu,\,\nu)$ are space-time indices, and $\alpha,\,\beta$ are the neutrino-flavor indices. The CPT-violating LIV parameter changes its sign under CPT transformation while the CPT-conserving one does not (see Refs.~\cite{Sahoo:2021dit, Kostelecky:2003cr}). Now, considering a realistic scenario where the neutrinos can propagate through the Earth matter, the effective Hamiltonian of an ultra-relativistic left-handed neutrino, in a three neutrino mixing scenario, can be expressed in the flavor basis as ~\cite{Kostelecky:2011gq, Sahoo:2021dit, Sahoo:2022rns}:
\begin{align}
	\mathcal{H}_{\rm eff} & = \frac{1}{2E}\,U\,\Delta m^2\,U^\dagger
	+ \frac{1}{E}\big(a^\mu_{L} p_\mu
	- c^{\mu\nu}_{L} p_\mu p_\nu \big) + \sqrt{2}G_FN_e\tilde{I}.
	\label{Eq:LIV-2}
\end{align} 
The first term in the above equation, $U$ represents the three-neutrino unitary mixing matrix, also known as PMNS matrix, and the $\Delta m^2$ part contains two independent mass-squared splittings in the form of a diagonal-matrix: ${\rm diag}(0,\,\Delta{m}^2_{21},\,\Delta{m}^2_{31})$. The second term shows the strength of induced potential with left-handed neutrino due to LIV, and $p$ is the four-momenta. The third term defines the effective matter-potential induced due to the elastic charged-current scattering between $\nu_e$ and electron. Here, $G_F$ is an electro-weak coupling constant, also known as the Fermi constant, $N_e$ is the number density of the ambient electrons present in matter, and $\tilde{I}$ is a diagonal matrix of the form diag$(1,\,0,\,0)$. The scalar part of the last term can be parameterized with matter density as $\sqrt{2}G_FN_e$ $\approx$ $7.6\,\times 10^{-23}\cdot Y_e \cdot \rho\left(\rm g/cm^3\right)$ GeV, where $Y_e$ is the relative electron number density in the ambient matter having an average Earth-matter density $\rho$.

In this study, we only focus on the time-like component ($\mu,\,\nu\,=\,0$) of LIV parameters in an isotropic space-time coordinate. From here onwards, we will consider $(a^0_L)_{\alpha\beta} \equiv a_{\alpha\beta}$ and $(c^{00}_L)_{\alpha\beta} \equiv c_{\alpha\beta}$.  Using the Sun-centered celestial-equatorial coordinate (see Ref.~\cite{Kostelecky:2008ts}) as an approximated inertial frame of reference, the Eq. \ref{Eq:LIV-2} can be written in the following fashion:

\begin{align}
	\mathcal{H}_{\rm eff} = \; \frac{1}{2E} 
	U\left(\begin{array}{ccc}
		0 & 0 & 0                 \\
		0 & \Delta m^{2}_{21} & 0 \\
		0 & 0 & \Delta m^{2}_{31} \\
	\end{array}\right)U^{\dagger}
	+&\left(\begin{array}{ccc}
		a_{ee} & a_{e\mu} & a_{e\tau} \\
		a^*_{e\mu} & a_{\mu\mu} & a_{\mu\tau} \\
		a^*_{e\tau} & a^*_{\mu\tau} & a_{\tau\tau}
	\end{array} \right) \nonumber \\
	-&\frac{4}{3} E
	\left(
	\begin{array}{ccc}
		c_{ee} & c_{e\mu} & c_{e\tau} \\
		c^*_{e\mu} & c_{\mu\mu} & c_{\mu\tau} \\
		c^*_{e\tau} & c^*_{\mu\tau} & c_{\tau\tau}
	\end{array}
	\right) 
	+ \sqrt{2}G_{F}N_{e}
	\left(\begin{array}{ccc}
		1 & 0 & 0 \\ 
		0 & 0 & 0 \\ 
		0 & 0 & 0
	\end{array}\right).
	\label{Eq:LIV-3}
\end{align}
For the case of right-handed antineutrino $U \to U^*$, $a_{\alpha\beta} \to -a_{\alpha\beta}^*$, $c_{\alpha\beta} \to c_{\alpha\beta}^*$ and $\sqrt{2}G_FN_e \to -\sqrt{2}G_FN_e$. Note that an extra fraction $4/3$ appears due to the choice of isotropic coordinates. It is essential to note that at the origin of the LIV parameters $a_{\alpha\beta}$ and $c_{\alpha\beta}$ are real-valued quantities~\cite{Colladay:1996iz}. However, due to the hermiticity, the off-diagonal elements of these LIV interaction matrices can have imaginary components while implanting them in an effective Hamiltonian.

To have an estimate of the strength of LIV parameters which may affect the outcome of the long-baseline experiments under consideration, we compare the first three terms in the neutrino propagation Hamiltonian as shown in Eq.~\ref{Eq:LIV-3}. The first term governs the neutrino oscillation in vacuum, whereas the second and third terms are the contribution from CPT-violating and CPT-conserving LIV, respectively. For typical long-baseline experiments with neutrino energy in the GeV range, the relevant scale for standard atmospheric neutrino oscillation is $\Delta m^2_{31}/{2E}\sim 10^{-22}$ GeV. So, in order to have noticeable effects from LIV, the strength of the parameters in the second and third terms should be around the same order as the standard neutrino oscillation scale. So, we observe that both CPT-violating ($a_{\alpha\beta}$) and CPT-conserving ($E\times c_{\alpha\beta}$) LIV parameters should have the strength of the order $10^{-22}$ GeV to have visible effects over standard neutrino oscillation in the matter.

\begin{table}[h!]
	\centering
	\begin{center}
		\begin{adjustbox}{width=1\textwidth}
			\begin{tabular}{|c| c| c| c|}
				\hline \hline
				\multicolumn{4}{|c|}{Existing constraints on CPT-violating LIV parameters} \\ \hline
				Experiments & $a_{e\mu} ~[ \rm GeV ]$ &  $a_{e\tau} ~[ \rm GeV ]$ & $a_{\mu\tau} ~[ \rm GeV ]$ \\ \hline
				\multirow{2}{*}{Super-K (95\% C.L.)} & $\mathrm{Re}(a_{e\mu}) < 1.8\times10^{-23}$  & $\mathrm{Re}(a_{e\tau}) < 4.1\times10^{-23}$ & $\mathrm{Re}(a_{\mu\tau}) < 0.65\times10^{-23}$ \\
				
				& $\mathrm{Im}(a_{e\mu}) < 1.8\times10^{-23}$  & $\mathrm{Im}(a_{e\tau}) < 2.8\times10^{-23}$ & $\mathrm{Im}(a_{\mu\tau}) < 0.51\times10^{-23}$ \\
				\hline
				\multirow{2}{*}{IceCube (99\% C.L.)} &\multirow{2}{*}{--} & \multirow{2}{*}{--} & $|\mathrm{Re}(a_{\mu\tau})| < 0.29\times10^{-23}$ \\ 
				&\multirow{2}{*}{--} & & $|\mathrm{Im}(a_{\mu\tau})| < 0.29\times10^{-23}$ \\ 
				\hline 
				\hline
				\multicolumn{4}{|c|}{Existing constraints on CPT-conserving LIV parameters} \\ \hline 
				
				Experiments & $c_{e\mu}$ &  $c_{e\tau}$ & $c_{\mu\tau}$ \\ \hline
				
				\multirow{2}{*}{Super-K (95\% C.L.)} & $\mathrm{Re}(c_{e\mu}) < 8.0\times10^{-27}$  & $\mathrm{Re}(c_{e\tau}) < 9.3\times10^{-25}$ & $\mathrm{Re}(c_{\mu\tau}) < 4.4\times10^{-27}$ \\
				
				& $\mathrm{Im}(c_{e\mu}) < 8.0\times10^{-27}$  & $\mathrm{Im}(c_{e\tau}) < 1.0\times10^{-24}$ & $\mathrm{Im}(c_{\mu\tau}) < 4.2\times10^{-27}$ \\
				
				\hline
				
				\multirow{2}{*}{IceCube (99\% C.L.)} &\multirow{2}{*}{--} & \multirow{2}{*}{--} & $|\mathrm{Re}(c_{\mu\tau})| < 0.39\times10^{-27}$ \\ 
				&\multirow{2}{*}{--} & & $|\mathrm{Im}(c_{\mu\tau})| < 0.39\times10^{-27}$
				\\\hline\hline
				
			\end{tabular}
		\end{adjustbox}
	\end{center}
	\mycaption{Existing constraints on the off-diagonal CPT-violating and CPT-conserving LIV parameters from Super-K~\cite{Super-Kamiokande:2014exs} and IceCube~\cite{IceCube:2017qyp}.}
	\label{tab:existing_bounds}
\end{table}

Following the above-discussed formalism, various neutrino experiments have given bounds on the CPT-violating and CPT-conserving LIV parameters. In particular, a recent publication by the IceCube collaboration~\cite{IceCube:2017qyp}, where the analysis has been performed in an effective two-flavor oscillation scenario, presented the most stringent bounds on the CPT-violating and the CPT-conserving LIV parameters in $\mu-\tau$ sector. Apart from this, there are also limits on both CPT-violating and CPT-conserving LIV parameters using the atmospheric neutrino data\footnote{Though the main focus of this paper is to explore the potential of DUNE and Hyper-K with their long-baseline setups, it is important to mention that these experiments will also collect huge atmospheric data which can be used to probe the LIV parameters under consideration in this paper. For instance, in Ref.~\cite{Hyper-Kamiokande:2018ofw}, the Hyper-K collaboration has already estimated the bounds on both CPT-violating and CPT-conserving LIV parameters which are 3 to 4 times more stringent than Super-K. This improvement is attributed to the larger fiducial volume of Hyper-K than Super-K. The DUNE collaboration has derived future limits on the LIV parameters associated with the Lorentz-violating operators of higher dimension (mass dimension $>$ 4) in Ref.~\cite{DUNE:2020ypp} with its atmospheric sample.} from Super-K~\cite{Super-Kamiokande:2014exs}. In Table~\ref{tab:existing_bounds}, we tabulate the existing limits on various off-diagonal LIV parameters from these two experiments. Note that in our work, we represent the off-diagonal LIV parameters as $|R|\cdot e^{i\phi}$, where $|R|$ denotes the magnitude of the LIV parameter and $\phi$ is the complex phase associated with it.

Several existing studies, for example, T2K near detector~\cite{Abe:2017eot}, MiniBooNE~\cite{MiniBooNE:2011pix}, MINOS near detector~\cite{MINOS:2008fnv}, MINOS far detector~\cite{MINOS:2010kat}, explore the possibility of sidereal time variation of neutrino flavor transition, considering the rotation of Earth in the LIV background. But these studies did not get any signature of the sidereal time dependence of neutrino oscillation probability and put limits on various combinations of LIV parameters. The sidereal dependence of the transition probability requires the presence of the spatial anisotropic LIV parameters~\cite{Diaz:2009qk}. However, since in the present work, we consider only the time-like components of the LIV parameters (one-at-a-time) in an isotropic coordinate system, the limits obtained from our work will not be directly applicable to the parameters that are probed in the above-mentioned studies.

\subsection{Analytical Expressions of the Oscillation Probabilities with LIV}

The presence of LIV terms in the neutrino Hamiltonian would affect the neutrino propagation through a medium, consequently modifying the neutrino flavor transition probability. So, it is possible to probe LIV in various neutrino oscillation experiments. In order to have an analytical understanding of the neutrino flavor transition probabilities in the presence of LIV, we follow the approach given in Refs.~\cite{Kikuchi:2008vq, Agarwalla:2016fkh, KumarAgarwalla:2019gdj}, where authors use perturbation theory to calculate the neutrino evolution matrix in various BSM scenarios like the presence of neutral current NSI and CPT-violating LIV parameters. We use $\alpha~(\equiv \Delta m^2_{21}/\Delta m^2_{31})$, $\sin^2\theta_{13}$, and LIV parameters $a'_{\alpha\beta}~(\equiv a_{\alpha\beta}/\sqrt{2}G_F N_e)$ and $c'_{\alpha\beta}~(\equiv c_{\alpha\beta}E/\sqrt{2}G_F N_e)$ ($\alpha,\beta = e,\mu,\tau$) as the expansion parameters. In this work, we mainly probe the off-diagonal CPT-violating and CPT-conserving LIV parameters $a_{\alpha\beta}$ and $c_{\alpha\beta}$ ($\alpha,\beta=e,\mu,\tau;\,\alpha\neq\beta$), respectively.

\vspace{0.3cm}
$\bullet$  \textbf{$\nu_\mu\to\nu_e$ Appearance Channel:}
\vspace{0.3cm}

The expression for the $\nu_\mu\to\nu_e$ transition probability, considering terms up to first-order in the above mentioned expansion parameters, can be written as~\cite{KumarAgarwalla:2019gdj},
\begin{align}\label{eq:pme_liv}
	\centering
	P_{\mu e} \simeq P_{\mu e}(\text{SI}) + P_{\mu e}(a_{e\beta}/c_{e\beta}) + \mathcal{O}(\alpha^2, \alpha \sin^2\theta_{13},a'^2_{e\beta},c'^2_{e\beta},a'^2_{\mu\tau},c'^2_{\mu\tau}),\,\,\,\,\,\,\,\beta=\mu,\tau.
\end{align}
The first term in right-hand-side (RHS) is the standard $\nu_\mu\to\nu_e$ appearance probability in the absence of any new physics parameters,
\begin{align}\label{eq:p_si}
	P_{\mu e}(\text{SI}) 
	&\simeq \mathbb{X} + \mathbb{Y}\cos(\delta_{\text{CP}} + \Delta),
\end{align}
where,
\begin{align}\label{eq:si_coeff}
	&\mathbb{X} = \sin^22\theta_{13}\sin^2\theta_{23} \frac{\sin^{2}\big[(1-\hat{A})\Delta \big]}{(1-\hat{A})^{2}};\nonumber \\
	&\mathbb{Y} = \alpha \sin2\theta_{12} \sin2\theta_{13} \sin2\theta_{23} \frac{\sin \hat{A}\Delta}{\hat{A}}  \frac{\sin \big[(1-\hat{A})\Delta\big]}{1-\hat{A}}; \nonumber \\
	&\hat{A} = \frac{2\sqrt{2}G_{F}N_{e}E}{\ldm}, \qquad  \Delta = \frac{\ldm L}{4E}.
\end{align}
The second term in RHS of Eq.~\ref{eq:pme_liv} are the contributions from the LIV parameters $a_{e\beta}/c_{e\beta}$ ($\beta = \mu,\tau$). For CPT-violating LIV cases, the expression of this term is;
\begin{align}\label{eq:p_cptv_liv}
	&P_{\mu e}(a_{e\beta}) 
	\simeq 
		2|a_{e\beta}|L\sin\tym\sin2\tzm \sin\Delta\big[\mathbb{Z}_{e\beta}\sin(\delta_{\rm CP} + \varphi_{e\beta}) +
		\mathbb{W}_{e\beta}\cos(\delta_{\rm CP} + \varphi_{e\beta})\big],\nonumber\\
\end{align}
and for CPT-conserving case,
\begin{align}\label{eq:p_cptc_liv}
	&P_{\mu e}(c_{e\beta}) 
	\simeq\frac{-8}{3}|c_{e\beta}|EL\sin\tym\sin2\theta_{23}\sin\Delta \big[\mathbb{Z}_{e\beta}\sin(\delta_{\rm CP} + \varphi_{e\beta}) +
		\mathbb{W}_{e\beta}\cos(\delta_{\rm CP} + \varphi_{e\beta})\big],\nonumber\\
\end{align}
where,
\begin{align}\label{eq:liv_coeff}
	&\mathbb{Z}_{e\beta} = 
	\begin{cases}
		- c_{23}  \sin \Delta, & \text{if}\ \beta=\mu. \\
		s_{23}  \sin \Delta, & \text{if}\ \beta=\tau.
	\end{cases} \nonumber \\
	&\mathbb{W}_{e\beta} = 
	\begin{cases}
		c_{23}  \big(\frac{s_{23}^{2}\sin \Delta}{c_{23}^{2}\Delta} + \cos \Delta \big), & \text{if}\ \beta=\mu. \\
		s_{23}  \big(\frac{\sin \Delta}{\Delta} - \cos \Delta \big), & \text{if}\ \beta=\tau.
	\end{cases} 
\end{align}
Note that the other off-diagonal LIV parameter $\amt/\cmt$ does not appear in the first-order terms. However, it may be present in higher-order terms and has a relatively smaller impact on the appearance probabilities. Note that for the appearance probability in the antineutrino case, one needs to apply $a_{\alpha\beta}\rightarrow-a^{*}_{\alpha\beta}$, $c_{\alpha\beta} \to c_{\alpha\beta}^*$, and $\hat{A}\to -\hat{A}$ in Eqs.~[\ref{eq:p_si} -\ref{eq:liv_coeff}]. In Appendix~\ref{appndx}, we show the validity of the approximate analytical expression of the appearance probability derived in this section by comparing it with the exact probability calculated numerically.

\begin{figure}[h!]
	\centering
	\includegraphics[width=\textwidth]{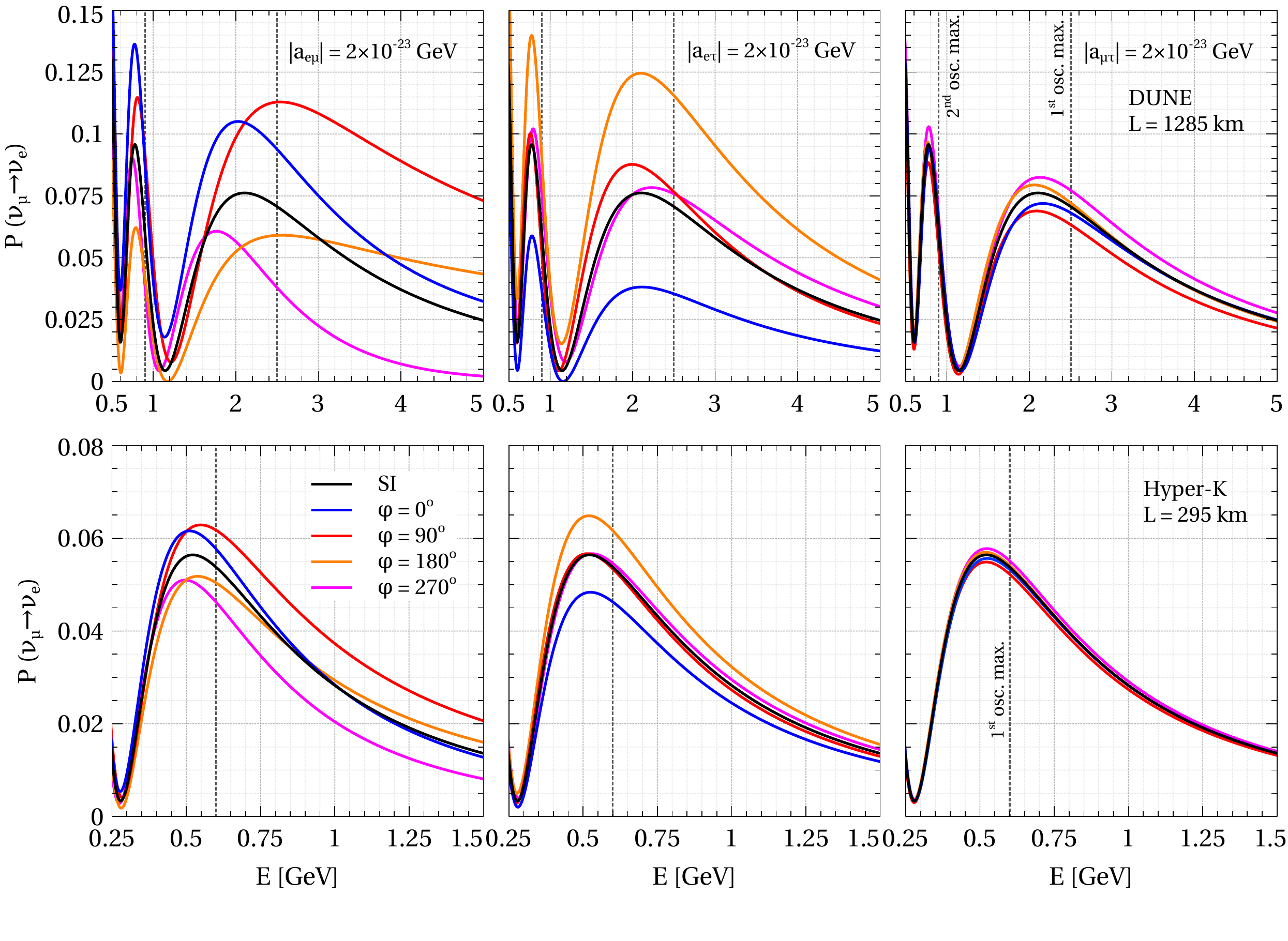}
	\vspace*{-10mm}
	\mycaption{$\nu_\mu\rightarrow\nu_e$ appearance probability as a function of energy in the presence of off-diagonal CPT-violating LIV parameters $a_{e\mu}$ (left column), $a_{e\tau}$ (middle column), and $a_{\mu\tau}$ (right column). The top and bottom rows correspond to the baselines of DUNE ($L=1285$ km) and Hyper-K ($L=295$ km), respectively. The black line in each panel shows the SI case (no LIV) and four colored lines correspond to four benchmark values of the phases associated with the LIV parameters: $0^\circ$, $90^\circ$, $180^\circ$, and $270^\circ$ with LIV strength $|a_{\alpha\beta}|=2.0\times10^{-23}$ GeV ($\alpha,\beta=e,\mu,\tau;\alpha\neq\beta$). The vertical grey-dashed lines in each panel show the energies at the first and second oscillation maxima. The values of the standard oscillation parameters used in this plot are mentioned in Table~\ref{tab:params_value}.}
	\label{fig:app_prob}
\end{figure}

\begin{table}[htb!]
	\centering
	 \begin{adjustbox}{width=0.7\textwidth}
	\begin{tabular}{|c|c|c|c|c|c|c|}
		\hline
		\hline
		$\theta_{12}$ & $\theta_{13}$ & $\theta_{23}$ & $\delta_{\text{CP}}$ & $\Delta{m^2_{21}}~[\rm{eV}^2]$ & $\Delta{m^2_{31}}~[\rm{eV}^2]$ & Mass Ordering\\
		\hline
		$33.45^\circ$ & $8.62^\circ$ & $42.1^\circ$ & $230^\circ$ & $7.42\times10^{-5}$ & $2.51\times10^{-3}$ & NMO\\
		\hline
		\hline
	\end{tabular}
   \end{adjustbox}
	\caption{Benchmark values of the oscillation parameters used in our analysis~\cite{Esteban:2020cvm}. We consider normal mass ordering (NMO) throughout this work which corresponds to $m_1<m_2<m_3$.}
	\label{tab:params_value}
\end{table}

\begin{figure}[h!]
	\centering
	\includegraphics[width=\textwidth]{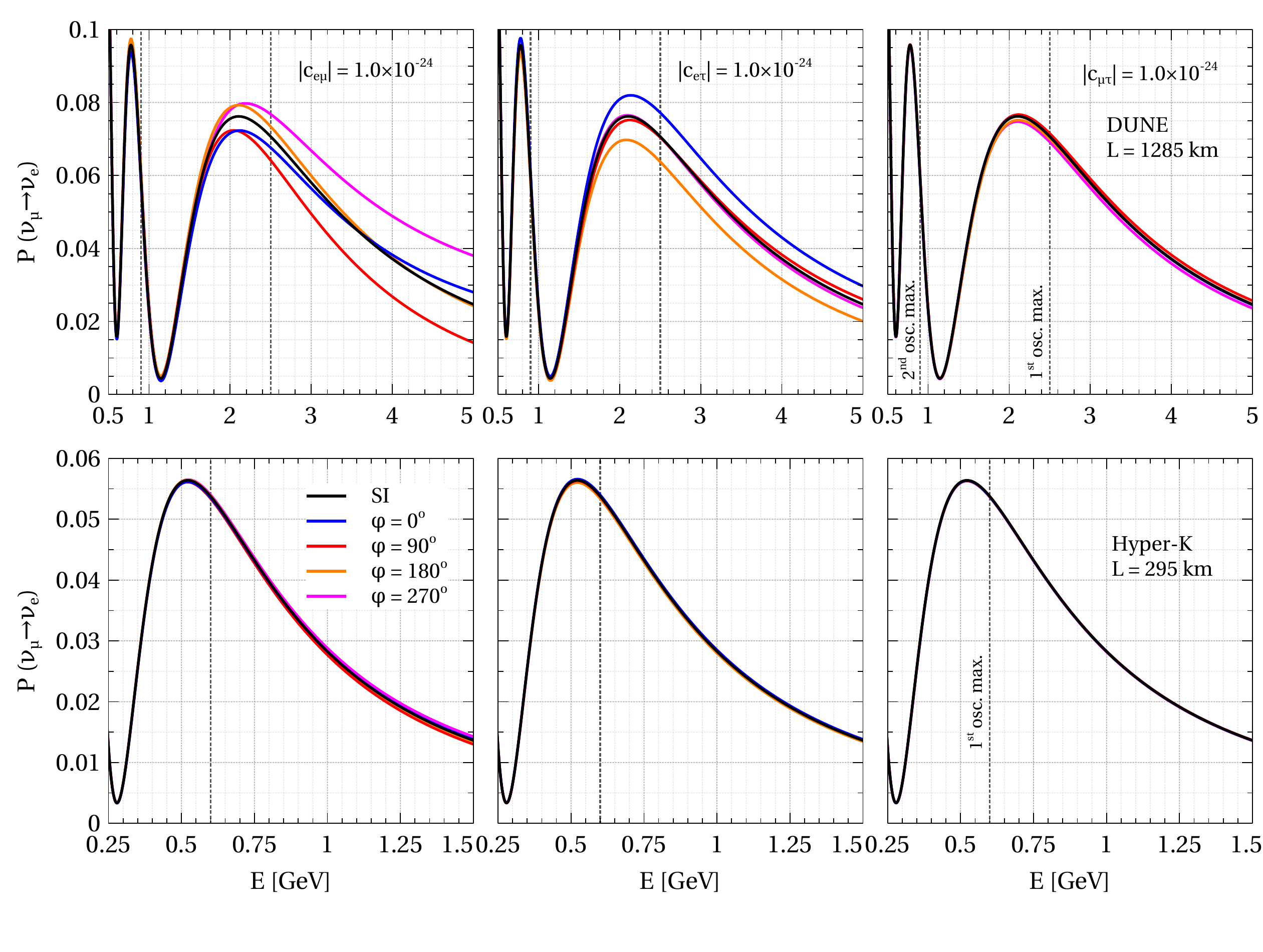}
	\vspace*{-10mm}
	\mycaption{$\nu_\mu\rightarrow\nu_e$ appearance probability as a function of energy in the presence of off-diagonal CPT-conserving LIV parameters $c_{e\mu}$ (left column), $c_{e\tau}$ (middle column), and $c_{\mu\tau}$ (right column). The top and bottom rows correspond to the baselines of DUNE ($L=1285$ km) and Hyper-K ($L=295$ km), respectively. The black line in each panel shows the SI case (no LIV) and four colored lines correspond to four benchmark values of the phases associated with the LIV parameters: $0^\circ$, $90^\circ$, $180^\circ$, and $270^\circ$ with LIV strength $|c_{\alpha\beta}|=1.0\times10^{-24}$ ($\alpha,\beta=e,\mu,\tau;\alpha\neq\beta$). The vertical grey-dashed lines in each panel show the energies at first and second oscillation maxima. The values of the standard oscillation parameters used in this plot are mentioned in Table~\ref{tab:params_value}.}
	\label{fig:cptc_app_prob}
\end{figure}

In Fig.~\ref{fig:app_prob}, we plot $\nu_\mu\rightarrow\nu_e$ oscillation probability as a function of neutrino energy for the baseline $L=1285$ km (top row) and $L=295$ km (bottom row) in the standard interaction (SI) case where there is no LIV and in the presence of the CPT-violating LIV parameters. To plot exact oscillation probability, we use GLoBES software~\cite{Huber:2004ka, Huber:2007ji} and modify the probability calculator accordingly to introduce LIV. The values of the standard oscillation parameters used to calculate the probability are given in Table~\ref{tab:params_value}. The left, middle, and right columns correspond to the appearance probability in the presence of  $a_{e\mu}$, $a_{e\tau}$, and $a_{\mu\tau}$ one-at-a-time, respectively with strength $|a_{\alpha\beta}|=2\times10^{-23}$ GeV. In each panel, the solid black curve shows the SI case. Four colored curves correspond to the probabilities in the presence of CPT-violating LIV parameter for the four chosen values of the associated phase, namely, 0, $90^\circ$, $180^\circ$, and $270^\circ$. It is clear from the figure that the impact of $\amt$ is marginal compared to $\aem$ and $\aet$. This behavior is obvious from our analytical expression in Eq.~\ref{eq:pme_liv} where $\amt$ does not appear at the leading order, contrary to the other two off-diagonal CPT-violating LIV parameters. In the presence of $\aem$, the appearance probability shows a significant deviation from the SI case depending on the value of the associated phase. Near first oscillation maxima, probability is maximum when $\phi_{e\mu}=90^\circ$ and minimum when $\phi_{e\mu}=270^\circ$. It can be explained using Eq.~\ref{eq:p_cptv_liv}, which shows the contribution from the CPT-violating LIV parameters to the oscillation probability. In case of $\aem$, the sign of $\mathbb{Z}$ ($\mathbb{W}$) in Eq.~\ref{eq:p_cptv_liv} is negative (positive) near the first oscillation maxima. Therefore, around the first oscillation maxima, the appearance probability is maximum (minimum) when the multiplicative factors associated with $\mathbb{Z}$ and $\mathbb{W}$ are negative (positive) and positive (negative), respectively. Since $\delta_{\rm CP}=230^\circ$, this happens at $\phi_{e\mu} = 90^\circ$ ($270^\circ$). However, in the middle panels, we observe that in the presence of $\aet$, the appearance probability is maximum (minimum) at $\phi_{e\tau}=180^\circ$ ($0^\circ$). This happens because now both $\mathbb{Z}$ and $\mathbb{W}$ are positive. So, the maximum (minimum) probability corresponds to the value of $\phi_{e\tau}$ for which both $\sin(\delta_{\rm CP} + \varphi_{e\beta})$ and $\cos(\delta_{\rm CP} + \varphi_{e\beta})$ in Eq.~\ref{eq:p_cptv_liv} are positive (negative). This occurs at $\phi_{e\tau} = 180^\circ$ ($0^\circ$) considering our benchmark value of $\delta_{\text{CP}}$.
Even though all these features can be observed both in the top and the bottom rows, we notice that the spread of the oscillation probability due to phase is significantly lesser in the bottom row, where we consider a comparatively smaller baseline ($L=295$ km). It is because of $L$ dependency in Eq.~\ref{eq:p_cptv_liv}, which adds the contribution from LIV parameters.

In Fig.~\ref{fig:cptc_app_prob}, we plot the appearance probability in the presence of three off-diagonal CPT-conserving LIV parameters, $\cem$ (left column), $\cet$ (middle column), and $\cmt$ (right column) with strength of $1.0\times10^{-24}$.
Here also, we use $L=1285$ km (top row) and $L=295$ km (bottom row). For DUNE ($L=1285$ km), in the presence of $c_{e\beta}$ ($\beta=\mu,\tau$), we observe that the impact of the associated phases is in the opposite order compared to the corresponding CPT-violating LIV parameters in Fig.~\ref{fig:app_prob}. It is because of the opposite sign of the LIV contributing term in the case of CPT-conserving LIV, as shown in Eq.~\ref{eq:p_cptc_liv}. As expected, $c_{\mu\tau}$ has almost a negligible effect on the probability in the case of DUNE. However, for Hyper-K ($L=295$ km), all three off-diagonal CPT-conserving LIV parameters have almost no impact on the oscillation probabilities because of the smaller baseline and comparatively lower neutrino energy.

\begin{figure}[h!]
	\centering	
	\includegraphics[width=\textwidth]{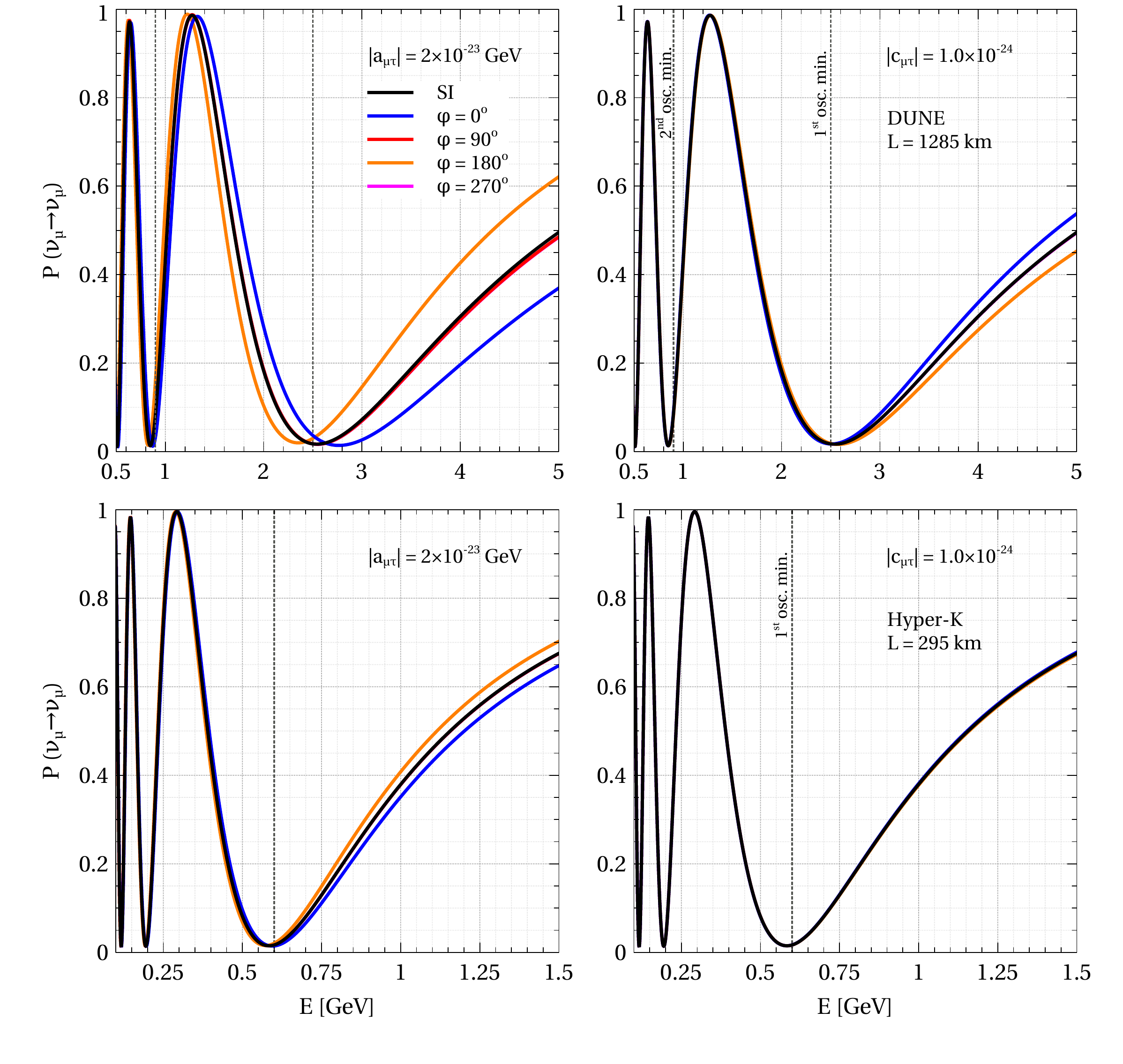}
	\vspace*{-10mm}
	\mycaption{$\nu_\mu\rightarrow\nu_\mu$ disappearance probability as a function of energy in the presence of off-diagonal LIV parameters $a_{\mu\tau}$ (left column), $c_{\mu\tau}$ (right column). The top and bottom rows correspond to the baselines of DUNE ($L=1285$ km) and Hyper-K ($L=295$ km), respectively. The black line in each panel shows the SI case (no LIV) and four colored lines correspond to four benchmark values of the phases associated with the LIV parameters: $0^\circ$, $90^\circ$, $180^\circ$, and $270^\circ$ with LIV strength $|a_{\alpha\beta}|=2.0\times10^{-23}$ GeV, $|c_{\alpha\beta}|=1.0\times10^{-24}$ ($\alpha,\beta=e,\mu,\tau;\alpha\neq\beta$). The vertical grey-dashed lines in each panel show the energies at first and second oscillation maxima. The values of the standard oscillation parameters used in this plot are mentioned in Table~\ref{tab:params_value}.}
	\label{fig:disapp_prob}
\end{figure} 

\vspace{0.25cm}
$\bullet$  \textbf{$\nu_\mu\to\nu_\mu$ Disappearance Channel:}
\vspace{0.25cm}

Now we discuss the impact of the CPT-violating and the CPT-conserving LIV parameters on the $\nu_\mu\to\nu_\mu$ disappearance probability.
Following the same strategy as the appearance channel, we derive the compact analytical expression for $\nu_\mu\to\nu_\mu$ disappearance probability,
\begin{align}\label{eq:pmm}
	P_{\mu\mu} \simeq P_{\mu\mu}(\text{SI}) + P_{\mu\mu}(\amt/\cmt) +\mathcal{O}(\alpha^2, \alpha \sin^2\theta_{13},a'^2_{e\beta},c'^2_{e\beta},a'^2_{\mu\tau},c'^2_{\mu\tau}),\,\,\,\,\,\,\,\beta=\mu,\tau.
\end{align}
The first term in the RHS is the standard disappearance probability without any new physics contribution,
\begin{align}
	P_{\mu\mu}(\text{SI}) = P_{\mu\mu}(\text{vacuum},\text{two flavor})+\alpha\mathbb{P}+\sin^2\theta_{13}\mathbb{Q}+\alpha\sin\theta_{13}\mathbb{R},
\end{align}
where,
\begin{align}
	P_{\mu\mu}(\text{vacuum},\text{two flavor}) = 1-\sin^22\theta_{23}\sin^2\Delta.
\end{align}
$\mathbb{P}$, $\mathbb{Q}$, and $\mathbb{R}$ are defined as follows,
\begin{align}
	&\mathbb{P} = \cos^2\theta_{12}\sin^22\theta_{23}\sin2\Delta,\\
	&\mathbb{Q} = -4\sin^2\theta_{23}\frac{\sin^2(\hat{A}-1)\Delta}{(\hat{A}-1)^2}-\frac{2}{\hat{A}-1}\sin^22\tzm\left(\sin\Delta\cos \hat{A}\Delta\frac{\sin(\hat{A}-1)\Delta}{\hat{A}-1}-\frac{\hat{A}}{2}\Delta\sin2\Delta\right),\\
	&\mathbb{R}= 2\sin2\theta_{12}\sin2\theta_{23}\cos\delta_{\rm CP}\cos\Delta\frac{\sin \hat{A}\Delta}{\hat{A}}\frac{\sin(\hat{A}-1)\Delta}{\hat{A}-1}.
\end{align}
Note that in the above expressions, we have also neglected the term with $\alpha\sin\tym\cos2\tzm$, since it is of the same order as $\alpha \sin^2\theta_{13}$. The contribution from the LIV parameters\footnote{Only off-diagonal LIV parameter that appears at the first order in the disappearance probability is $\amt/\cmt$. This has already been discussed in Refs.~\cite{Kopp:2007ne,Kikuchi:2008vq} in case of NSI.} is given by,
\begin{align}\label{eq:Pmm_liv}
	P_{\mu\mu}(\amt/\cmt) = \frac{\sin^22\tzm}{2}\left[2\sin^2\tym\Delta-\mathbb{S}\right] \sin2\Delta,
\end{align}
where,
\begin{align}\label{eq:S}
	\mathbb{S} = \begin{cases}
		2L\sin2\tzm|a_{\mu\tau}|\cos\phi_{\mu\tau}, & \text{CPT-violating LIV}. \\
		-\frac{8}{3}E L \sin2\tzm|c_{\mu\tau}|\cos\phi_{\mu\tau}, & \text{CPT-conserving LIV}.
	\end{cases} 
\end{align}
In Appendix~\ref{appndx}, we compare the disappearance probability calculated using the analytical expressions derived above with the same calculated numerically.

In Fig.~\ref{fig:disapp_prob}, we show the $\nu_\mu\to\nu_\mu$ disappearance probability as a function of energy for the baseline $L=1285$ km (top row) and $L=295$ km (bottom row) in SI case as well as in the presence of LIV parameters. In the left column, we show the impact of CPT-violating LIV parameter $\amt$, whereas the right column corresponds to the CPT-conserving LIV parameter $c_{\mu\tau}$. As discussed earlier, the other two off-diagonal parameters do not appear at the leading order and are expected to have a negligible impact on disappearance probability. From the left panels, we observe that the phase associated with $\amt$, shows a significant impact with positive (negative) deviation for $\phi_{\mu\tau}=180^\circ$ ($0^\circ$) from the SI case. We can explain this feature using the LIV contributing terms in the analytical expression (ref. to Eq.~\ref{eq:pmm}). It is clear that when $\mathbb{S}$ (shown in Eq.~\ref{eq:S}) is negative (positive), the disappearance probability is larger (smaller) than its corresponding SI value. Similar to the appearance probability, here also, we observe the impact of the phases become smaller for a smaller baseline, as shown in lower panels, because the LIV contributing term is proportional to the baseline length $L$. In the right column, we show the impact of the CPT-conserving LIV parameters. Here, we observe that for $L=1285$ km, disappearance probability is maximum (minimum) for $\phi_{\mu\tau}=0^\circ$ ($180^\circ$), which shows an opposite behavior compared to the CPT-violating LIV parameters. This attributes to the opposite sign of $\mathbb{S}$ in Eq.~\ref{eq:S}, in the case of the CPT-conserving LIV parameters. 

\section{Long-baseline Experiments: DUNE and Hyper-K}
\label{sec:LBL}
\subsection{Essential Features of the Experimental Setups}
Accelelarator-based neutrino oscillation experiments are playing a very important role in resolving issues in the standard $3\nu$ paradigm and exploring various BSM physics in the neutrino sector. Precise information about the neutrino flux, cross section, and baseline make these experiments unique. In this work, we probe LIV in the context of next-generation long-baseline experiments DUNE and Hyper-K. DUNE is a future long-baseline experiment with an on-axis, high-intensity wide-band neutrino beam produced at Fermilab~\cite{DUNE:2020lwj, DUNE:2020jqi, DUNE:2021cuw, DUNE:2021mtg}. The detector would be a 40 kt Liquid argon time projection chamber (LArTPC) placed underground at Homestake mine, 1285 km from the source. On the other hand, Hyper-K~\cite{Abe:2015zbg, Hyper-Kamiokande:2018ofw} is another next-generation long-baseline experiment with off-axis, narrow band beam produced at the J-PARC proton synchrotron facility. The beam would be detected at Hyperkamiokande, a 187 kt water Cherenkov detector placed at a distance of 295 km from the source with an off-axis angle $2.5^\circ$. In Table~\ref{tab:exp_details}, we give the other relevant information on these two experiments.

\begin{table}
	\centering
 \begin{adjustbox}{width=0.8\textwidth}
	\begin{tabular}{|c|c|c|}
		\hline \hline
		& DUNE & Hyper-K\\ 
		\hline
		Detector Mass& 40 kt LArTPC & 187 kt WC   \\
		\hline
		Baseline & 1285 km & 295 km \\
		\hline
		Proton Energy & 120 GeV & 80 GeV \\
		\hline
		Beam type & Wide-band, on-axis & Narrow-band, off-axis ($2.5^{\circ}$)\\
		\hline
		Beam power & 1.2 MW & 1.3 MW \\
		\hline
		P.O.T./year& $1.1\times10^{21}$ & $2.7\times10^{21}$\\ 
		\hline
		Run time ($\nu+\bar{\nu}$) & 5 yrs + 5 yrs & 2.5 yrs + 7.5 yrs  \\
		\hline
		Normalization error & 2\% (app.) 5\% (disapp.) & 5\% (app.), 3.5\% (disapp.)\\
		\hline\hline
	\end{tabular}
 \end{adjustbox}
	\mycaption{Major features of LBL experiments, DUNE~\cite{DUNE:2020lwj} and Hyper-K~\cite{Abe:2015zbg} used in our simulation.}
	\label{tab:exp_details}
\end{table}

As mentioned earlier, the DUNE setup uses an on-axis, wide-band neutrino beam. This allows DUNE to explore both the first and second oscillation maxima for the baseline 1285 km, which are around 2.5 GeV and 0.9 GeV, respectively. On the other hand, Hyper-K has an off-axis narrow band beam with an energy peak around 0.6 GeV, which is the first oscillation maxima for Hyper-K. The monochromatic beam will give the advantage of high statistics at the first oscillation maxima, where the impact of various physics can be significant. 
DUNE will have equal runtime for neutrino and antineutrino mode, which will give a larger number of expected neutrino events than the antineutrino, since the neutrino has almost three times the cross section of the antineutrino. For Hyper-K, antineutrino runtime is three times the neutrino runtime in order to compensate for suppression in the cross section. So, depending on the physics under probe, different ratios of the neutrino and antineutrino events can be useful. The longer baseline of DUNE allows it to have larger matter effect compared to Hyper-K. Apart from the complementarity at the neutrino flux, baseline, and runtime, the detector properties of the two setups are different.
As proposed by the collaborations, DUNE is going to have total systematic uncertainties of 2.5\% in the appearance channel and 5\% in the disappearance channel~\cite{DUNE:2020ypp, DUNE:2020jqi, DUNE:2021cuw} and for Hyper-K, it is 5\% and 3.5\%~\cite{Hyper-KamiokandeWorkingGroup:2014czz, Abe:2015zbg}, respectively. Lower systematics in the appearance channel in DUNE can allow it to have comparatively larger sensitivity to some physics that have a larger impact on the appearance channel. A similar argument can be given for Hyper-K in the disappearance channel. Note that the mentioned values of the systematic uncertainties are estimated values. In future, these values may change, which would affect the results presented in the later sections.

Each oscillation channel in both the experiments have backgrounds. For DUNE, the appearance channel has background from intrinsic $\nu_e$ beam contamination and misidentified $\nu_\mu$, $\nu_\tau$, and neutral current (NC) events. For the disappearance channel, backgrounds are misidentified $\nu_\tau$ and NC events. Similarly, for Hyper-K, backgrounds in the appearance channel come from the $\nu_e$ beam contamination and misidentified $\nu_\mu$ and NC events. Backgrounds for the disappearance channel come from the misidentified $\nu_e$ and NC events.
\subsection{Expected Event Rates in the Presence of LIV}
\label{subsec:LBL-Evt-Sim}

As mentioned earlier, in this work, we consider two experimental configurations, DUNE and Hyper-K, for our analysis. We calculate the expected event rates of these two configurations using the GLoBES software~\cite{Huber:2004ka, Huber:2007ji}. For the oscillation analysis with LIV parameters, we use GLoBES-extension $snu.c$~\cite{Kopp:2007ne}.

In Table.~\ref{tab:total_events}, we give the expected $\nu_e/\bar{\nu}_e$ and $\nu_\mu/\bar{\nu}_\mu$ event rates from DUNE and Hyper-K in SI case and in the presence of various LIV parameters. Assumed configurations of the experiments are discussed in detail in Sec.~\ref{sec:LBL} (see Table~\ref{tab:exp_details}). While generating the events, the strength of the CPT-violating (CPT-conserving) LIV parameters is considered to be $2.0\times10^{-23}$ GeV ($1.0\times10^{-24}$), one at-a-time. Here, we consider the phase associated with the off-diagonal LIV parameters to be zero ($\phi_{\alpha\beta}=0^\circ$; $\alpha, \beta = e, \mu, \tau$ and $\alpha\neq\beta$) for a demonstration purpose.

	\begin{table}[h!]		
		\centering
		\begin{adjustbox}{width=1\textwidth}
			\begin{tabular}{|ccc|*{8}{c|}}		
				\hline\hline
				
				\multicolumn{3}{|c|}{\multirow{2}{*}{}} & \multicolumn{2}{|c}{$\nu_e$ appearance} & \multicolumn{2}{|c|}{$\bar\nu_e$ appearance}& \multicolumn{2}{|c}{$\nu_\mu$ disappearance} & \multicolumn{2}{|c|}{$\bar\nu_\mu$ disappearance}\\ 
				
				\cline{4-11}
				
				\multicolumn{3}{|c|}{} & DUNE & Hyper-K & DUNE & Hyper-K & DUNE & Hyper-K & DUNE & Hyper-K \\
				
				\hline
				
				\multicolumn{3}{|c|}{SI} & 1614 & 1613  & 292 & 727 & 15624 & 9577  & 9052 & 9074 \\
				
				\cline{1-11}
				
				\hline
				
				\multicolumn{3}{|c|}{$|a_{e\mu}|$ (= $2\times10^{-23}$ GeV)} & 2276 & 1731 & 666 & 901  & 15446 & 9581  & 8891 & 9042\\
				\hline
				\multicolumn{3}{|c|}{$|a_{e\tau}|$ (= $2\times10^{-23}$ GeV)} & 817 & 1387  & 226 & 697 & 15613 & 9565   & 9063 & 9084\\
				\hline
				\multicolumn{3}{|c|}{$|a_{\mu\tau}|$ (= $2\times10^{-23}$ GeV)} & 1567 & 1598  & 303 & 735 & 14404 & 9366  & 9237 & 9373\\
				\hline
				\multicolumn{3}{|c|}{$|c_{e\mu}|$ (= $1.0\times10^{-24}$)} & 1757 & 1610 & 480 & 735  & 15315 & 9575  & 8800 & 9071\\
				\hline
				\multicolumn{3}{|c|}{$|c_{e\tau}|$ (= $1.0\times10^{-24}$)} & 1792 & 1623  & 296 & 724 & 15634 & 9578   & 9056 & 9074\\
				\hline
				\multicolumn{3}{|c|}{$|c_{\mu\tau}|$ (= $1.0\times10^{-24}$)} & 1620 & 1614  & 295 & 727 & 16263 & 9600  & 9411 & 9099\\
				
				\hline
				\hline
			\end{tabular}
		\end{adjustbox}
		\mycaption{Total signal rate for the $\nu_{e}$ appearance channel and  $\nu_{\mu}$ disappearance channel both in neutrino and antineutrino mode for DUNE and Hyper-K setups in SI case as well as in the presence off-diagonal LIV parameters. The relevant features of these facilities are given in Table~\ref{tab:exp_details}. The strength of the CPT-violating (CPT-conserving) LIV parameters is taken to be $2\times10^{-23}$ GeV ($1.0\times10^{-24}$). The phases associated with the off-diagonal LIV parameters are considered to be zero. The values of the standard oscillation parameters used to calculate event rate are quoted in Table~\ref{tab:params_value}.}
	
	\label{tab:total_events}
\end{table}

We make following observations from Table~\ref{tab:total_events}:
\begin{itemize}

\item In the presence of $a_{e\mu}$ ($a_{e\tau}$), $\nu_e$ event deviates from the SI case by 41\% (49\%) for DUNE and by 7.3\% (14\%) for Hyper-K. 

\item Similarly, we observe that the presence of LIV parameter $a_{\mu\tau}$ changes the expected $\nu_{\mu}$ disappearance event rates by 7.8\% for DUNE and 2.2\% for Hyper-K from the SI case. In the presence of other LIV parameters, changes in the event rates are very small ($\leq 1\%$).

\item In presence of $c_{e\mu}$ ($c_{e\tau}$), the $\nu_{e}$ appearance event rates changes by $\sim$9\% (11\%) for DUNE, but for Hyper-K, the changes are very minute ($<$ 1\%). However $c_{\mu\tau}$ changes the $\nu_{\mu}$ event rates by 4\% for DUNE and 0.2\% for Hyper-K.

\end{itemize}

All these observations are consistent with the results seen at the probability level in the previous section. Though, in this section, we give only the total signal event rates for DUNE and Hyper-K, however, we perform a binned study while presenting the sensitivity results  in section~\ref{sec:results}. For demonstration purpose, we show in appendix~\ref{appndx-B}, the $\nu_e$ appearance and $\nu_\mu$ disappearance signal event spectra expected to be seen by DUNE and Hyper-K in SI case as well as in presence of LIV (SI+LIV). Similar plots can be made for antineutrinos. Note that while calculating the final sensitivity results, we take bin contributions from appearance and disappearance channels in both neutrino and antineutrino modes.

\begin{table}[h!]
	\begin{center}
		\begin{tabular}{|c|c|c|c|}
			\hline\hline
			Experiments & $E_{\rm rec}$ (GeV)& Bin width (GeV) & Total bins \\
			\hline\hline
			DUNE & \makecell[c]{$0.5\,-\, 8$ \\ $8\,-\,10$ \\ $10\,-\,18$} & 
			\makecell[c]{$0.125$ \\ $1.0$ \\ $2.0$} & 
			$\left.\begin{tabular}{l}
			60 \\ 2 \\ 4
			\end{tabular}\right\}$  $66$ \\
			\hline
			Hyper-K & $0.1\,-\,3.0$ & 0.1 & 29\\
			\hline\hline
		\end{tabular}
		\mycaption{The binning scheme used in our analysis for both DUNE and Hyper-K. The binning is performed over the reconstructed energy ($E_{\rm rec}$).\label{tab:binning}}
	\end{center}
\end{table}

\section{Numerical Analysis}
\label{sec:RnA}
One of the major goals of this work is to study the ability of DUNE and Hyper-K to constrain the off-diagonal CPT-violating and CPT-conserving LIV parameters. To estimate the sensitivity of a given experiment towards the LIV parameters, we use the following form of the Poissonian $\chi^2$:
\begin{equation}
\chi^2 (\vec{\lambda}, \xi_{s}, \xi_{b}) = \min_{\vec\rho, \xi_{s}, \xi_{b}}~\Big[{{2\sum_{i=1}^{n}}\big(y_{i}-x_{i}-x_{i} \text{ln}\frac{y_{i}}{x_{i}}\big) + \xi_{s}^2 + \xi_{b}^2}~\Big],
\end{equation} 
which gives the median sensitivity of the experiment where $n$ is the total number of reconstructed energy bins. The binning scheme adopted in this work for the simulation of DUNE and Hyper-K are given in Table~\ref{tab:binning}.
\begin{equation}
y_{i} = N^{th}_{i}
(\vec\rho)~[1 + \pi^{s}\xi_{s}] + N^{b}_{i}(\vec\rho)~[1+\pi^b\xi_{b}],
\end{equation}
\noindent
where $N^{th}_{i}$ is the expected number of signal events in the $i$-th bin with the set of oscillation parameters $\vec\lambda$ = \{$\theta_{12}, \theta_{13}, \theta_{23}, \Delta{m}^2_{21}, \Delta{m}^2_{31}, \delta_{\rm CP}, a_{\alpha\beta}, \phi^{a}_{\alpha\beta}, c_{\alpha\beta}, \phi^{c}_{\alpha\beta}$\}. $\phi^{a}_{\alpha\beta}$ and $\phi^{c}_{\alpha\beta}$ are the phases associated with the off-diagonal CPT-violating and CPT-conserving LIV parameters, respectively. $N^{b}_{i}$ is the number of background events in the $i$-th energy bin. The systematic pulls on the signal and background are denoted by the variables $\xi_{s}$ and $\xi_{b}$, respectively. We marginalize the $\chi^2$ over the set of parameters $\vec\rho$ and also over the systematic pulls ($\xi_{s}$ and $\xi_{b}$) in the fit. The variables $\pi^s$ and $\pi^b$ stand for the normalization error on the signal and background. The values of the normalization errors for DUNE and Hyper-K are listed in Table~\ref{tab:norm_err}. $x_{i} = N^{obs}_{i} + N^{b}_{i}$ embodies the prospective data from the experiment, where $N^{obs}_{i}$ is the number of charged current (CC) signal events and $N^b_{i}$, as mentioned before, is the number of background events. 

We quantify our results in terms of the statistical significance given by Poissonian $\Delta{\chi}^2$ defined as,
\begin{equation}
	\Delta{\chi}^2 = \min_{\vec\rho, \xi_{s}, \xi_{b}}\Big[\chi^2 (a_{\alpha\beta}/c_{\alpha\beta} \neq 0) - \chi^2 (a_{\alpha\beta}/c_{\alpha\beta} = 0)\Big],\,\,\,\,\,\alpha,\beta=e,\mu,\tau;\alpha\neq\beta.
\end{equation}
The first term in the right-hand side of the above equation is obtained when we fit the prospective data from the experiment with the theory in the presence of LIV, and the second term is calculated by fitting with the standard case with no LIV present in theory. Due to suppression in the statistical fluctuations, we can take $\chi^2 (a_{\alpha\beta}/c_{\alpha\beta} = 0)$ $\sim$ 0 while obtaining the median sensitivity of the experiment using frequentist approach~\cite{Blennow:2013oma}. Here $\vec\rho$ is the set of parameters over which the $\chi^2$ is marginalized, and $\xi_{s}$ and $\xi_{b}$ are the systematic pulls on the signal and background, respectively.
For our analysis, we use the true values of the standard oscillation parameters as given in Table~\ref{tab:params_value}. In theory, we keep the two mixing angles $\theta_{12}$, $\theta_{13}$ and two mass-splittings $\Delta m^2_{21}$, $\Delta m^2_{31}$ fixed at the same values. We do not marginalize over the present uncertainty in the magnitude of $\Delta m^2_{31}$, because the global oscillation data attain a relative 1$\sigma$ precision of 1.1\% in the measurement of $\Delta m^2_{31}$. JUNO, with six years of data, may further improve the measurement of this parameter to 0.2\%~\cite{JUNO:2022mxj}. Also, we do not marginalize over the neutrino mass ordering since there is a hint towards the normal mass ordering from the global oscillation data~\cite{deSalas:2020pgw, Esteban:2020cvm, Capozzi:2021fjo}. Moreover, T2K~\cite{T2K:2023smv}, NO$\nu$A~\cite{Carceller:2023kdz}, Super-K~\cite{Posiadala-Zezula:2022vzn}, IceCube-DeepCore~\cite{Mead:2023spo}, and ORCA~\cite{KM3NeT:2023ncz} are going to collect more data in the coming days which will further strengthen the estimation of mass ordering. Also, it is expected that with preliminary data coming from $\sim$ 3 years run of DUNE would be able to settle the issue of neutrino mass ordering at very high confidence level~\cite{DUNE:2020ypp}. For the correlation between LIV parameters and $\delta_{\mathrm{CP}}$, we marginalize over the $\theta_{23}$ in its allowed $3\sigma$ range~\cite{Esteban:2020cvm} and the phase $\phi$ associated with the off-diagonal LIV parameters in the range [0, $360^\circ$]. Similarly, for the LIV parameter and $\theta_{23}$ correlation analysis, we marginalize over $\delta_{\mathrm{CP}}$ in its allowed $3\sigma$ range and $\phi$ in its entire permitted range.
While deriving the limits on the off-diagonal LIV parameters, we marginalize over $\theta_{23}\in[39.7^\circ:50.9^\circ]$, $\delta_{\rm CP}\in[144^\circ:350^\circ]$, and $\phi\in[0^\circ:360^\circ]$.

\begin{table}[t!]
	\resizebox{\columnwidth}{!}{%
		\centering
		\begin{tabular}{|c|c|c|c|c|c|c|c|c|}
			\hline\hline
			\multirow{3}{*}{Expts.}  & \multicolumn{8}{c|}{Normalization errors~[\%]}  \\
			\cline{2-9}
			& \multicolumn{4}{c|}{Signal ($\pi^s$)} & \multicolumn{4}{c|}{Background ($\pi^b$)}\\ \cline{2-9}
			&  $\nu$ App. & $\bar{\nu}$ App. & $\nu$ Disapp.& $\bar{\nu}$ Disapp. & $\nu_{e}$, $\bar{\nu}_{e}$ CC & $\nu_{\mu}$, $\bar{\nu}_{\mu}$ CC & $\nu_{\tau}$, $\bar{\nu}_{\tau}$ CC & NC \\ 
			\hline
			DUNE & 2 & 2 & 5 & 5 & 5 & 5 & 20 & 10\\
			Hyper-K & 5 & 5 & 3.5 & 3.5 & 10 & 10 & -- & 10\\
			\hline\hline
	\end{tabular}}
	\caption{The values of the normalization errors associated to the signal and background event rates in DUNE and Hyper-K for neutrino and antineutrino in both appearance and disappearance channels. These values are taken from Ref.~\cite{DUNE:2021cuw} for DUNE and Ref.~\cite{Hyper-Kamiokande:2018ofw} for Hyper-K.}
	\label{tab:norm_err}
\end{table}

\section{Our Results}
\label{sec:results}
In this section, we present our results in two parts. First, we discuss the correlations between the LIV parameters and the most unsettled standard oscillation parameters, $\delta_{\mathrm{CP}}$ and $\theta_{23}$. This helps us to find if there is any degeneracy between the LIV parameters and the standard oscillation parameters. In the second part, we present the expected constraints on the LIV parameters from DUNE, Hyper-K individually, and their combination DUNE+Hyper-K.

\subsection{Correlations in test ($\delta_{\rm CP} - |a_{\alpha\beta}|/|c_{\alpha\beta}|$) and test ($\theta_{23} - |a_{\alpha\beta}|/|c_{\alpha\beta}|$) Planes}
\label{subsec:correlation}

\begin{figure}[h!]
	\centering
	\includegraphics[width=\textwidth]{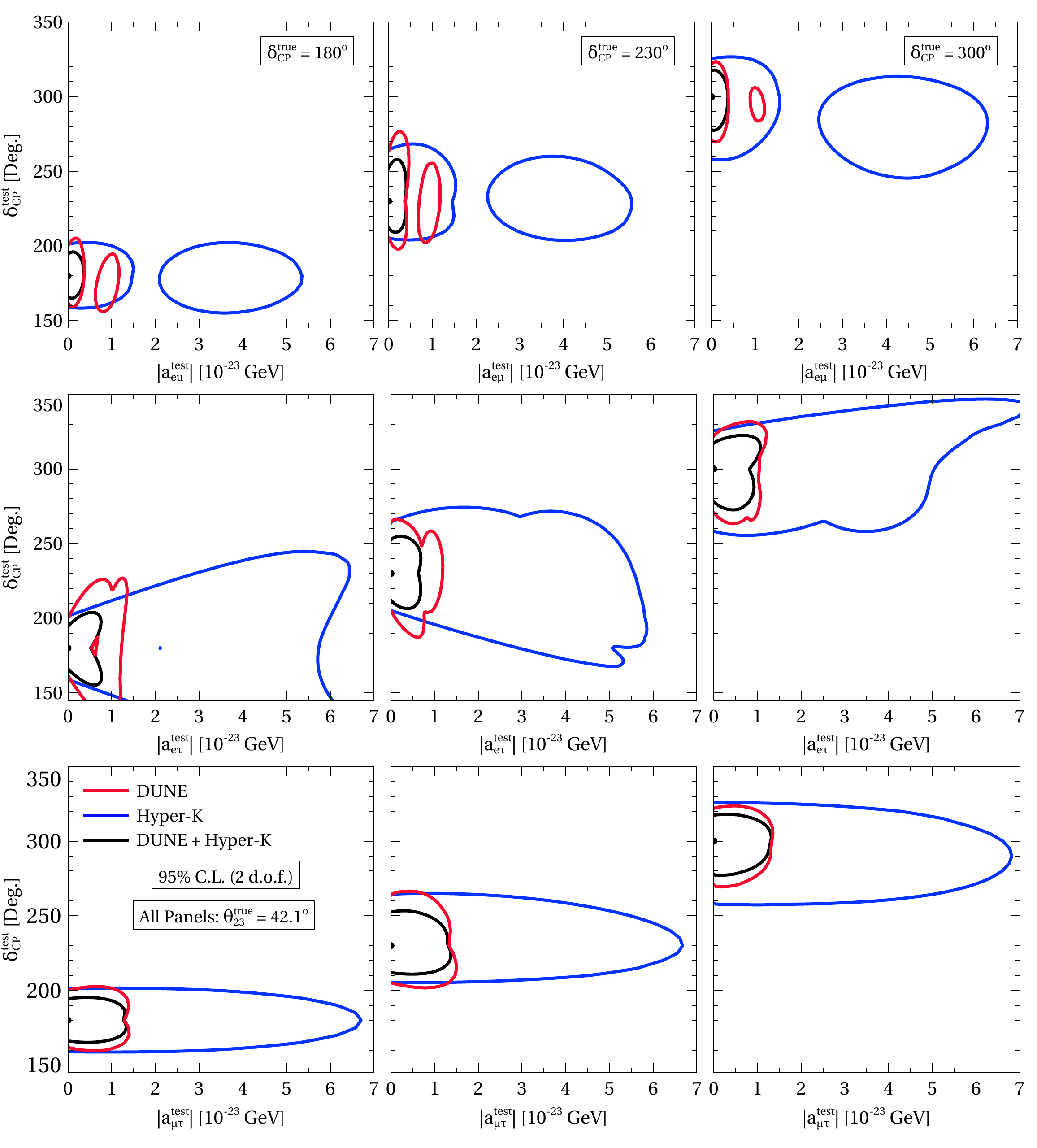}
	\vspace*{-10mm}
	\mycaption{95\% C.L. (2 d.o.f.) contours in the $\delta_{\rm CP}-|a_{e\mu}|$ (top row), $\delta_{\rm CP}-|a_{e\tau}|$ (middle row), and $\delta_{\rm CP}-|a_{\mu\tau}|$ (bottom row) planes for DUNE, Hyper-K, and DUNE+Hyper-K. Three benchmark values of $\delta_{\mathrm{CP}}$ considered in the data are $180^\circ$ (left column), $230^\circ$ (middle column), and $300^\circ$ (right column), as shown by the black dots in each panel. True values of other oscillation parameters are given in Table~\ref{tab:params_value}. In the fit, we marginalize over $\theta_{23}$ in its allowed $3\sigma$ range and new phase $\phi$ in its entire allowed range.}
	\label{fig:correlation_dcp_cptv}
\end{figure}

\begin{figure}[h!]
	\centering
	\includegraphics[width=\textwidth]{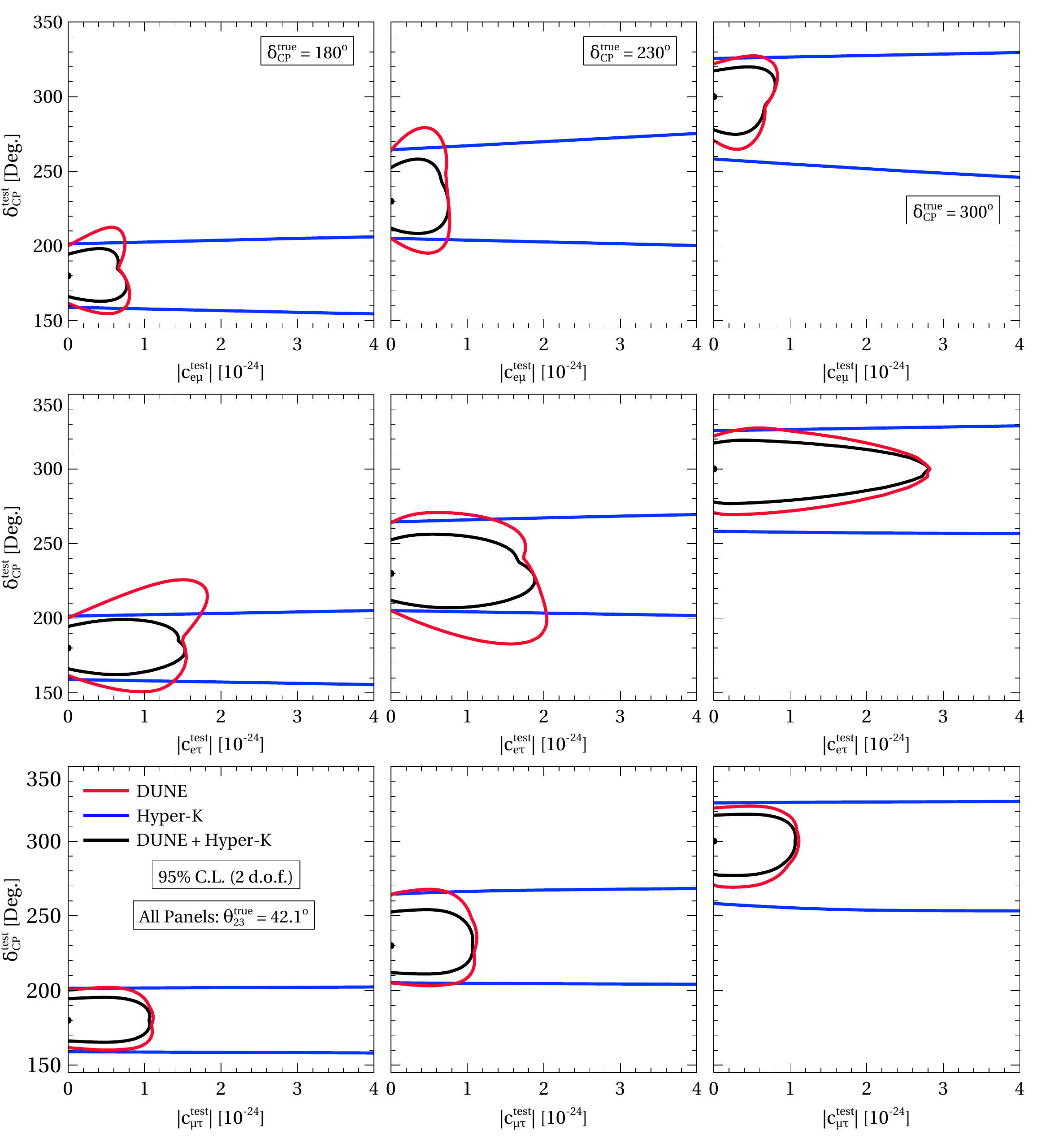}
	\vspace*{-10mm}
	\mycaption{95\% C.L. (2 d.o.f.) contours in the $\delta_{\rm CP}-|c_{e\mu}|$ (top row), $\delta_{\rm CP}-|c_{e\tau}|$ (middle row), and $\delta_{\rm CP}-|c_{\mu\tau}|$ (bottom row) planes for DUNE, Hyper-K, and DUNE+Hyper-K. Three benchmark values of $\delta_{\mathrm{CP}}$ considered in the data are $180^\circ$ (left column), $230^\circ$ (middle column), and $300^\circ$ (right column), as shown by black dots in each panel. True values of other oscillation parameters are given in Table~\ref{tab:params_value}. In the fit, we marginalize over $\theta_{23}$ in its allowed $3\sigma$ range and new phase $\phi$ in its entire allowed range.}
	\label{fig:correlation_dcp_cptc}
\end{figure}

\begin{figure}[h!]
	\centering
	\includegraphics[width=\textwidth]{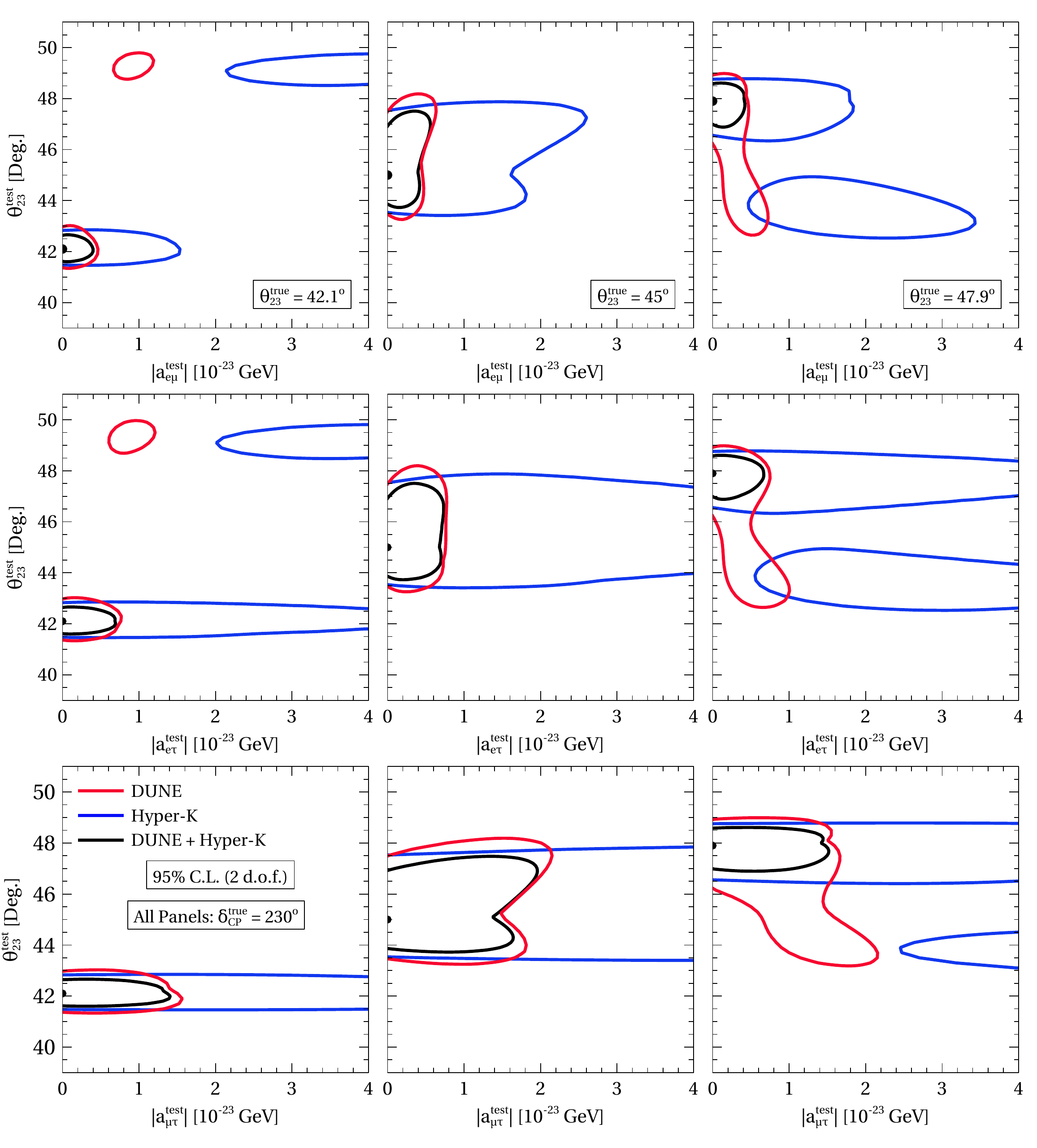}
	\vspace*{-10mm}
	\mycaption{ 95\% C.L. (2 d.o.f.) contours in the $\tzm-|a_{e\mu}|$ (top row), $\tzm-|a_{e\tau}|$ (middle row), and $\tzm-|a_{\mu\tau}|$ (bottom row) planes for DUNE, Hyper-K, and DUNE+Hyper-K. Three benchmark values of $\tzm$ considered in the data are $42.1^\circ$ (left column), $45^\circ$ (middle column), and $47.9^\circ$ (right column), as shown by black dots in each panel. True values of other oscillation parameters are given in Table~\ref{tab:params_value}. In the fit, we marginalize over $\delta_{\mathrm{CP}}$ in its allowed $3\sigma$ range and new phase $\phi$ in its entire allowed range.}
	\label{fig:correlation_th23_cptv}
\end{figure}

\begin{figure}[h!]
	\centering
	\includegraphics[width=\textwidth]{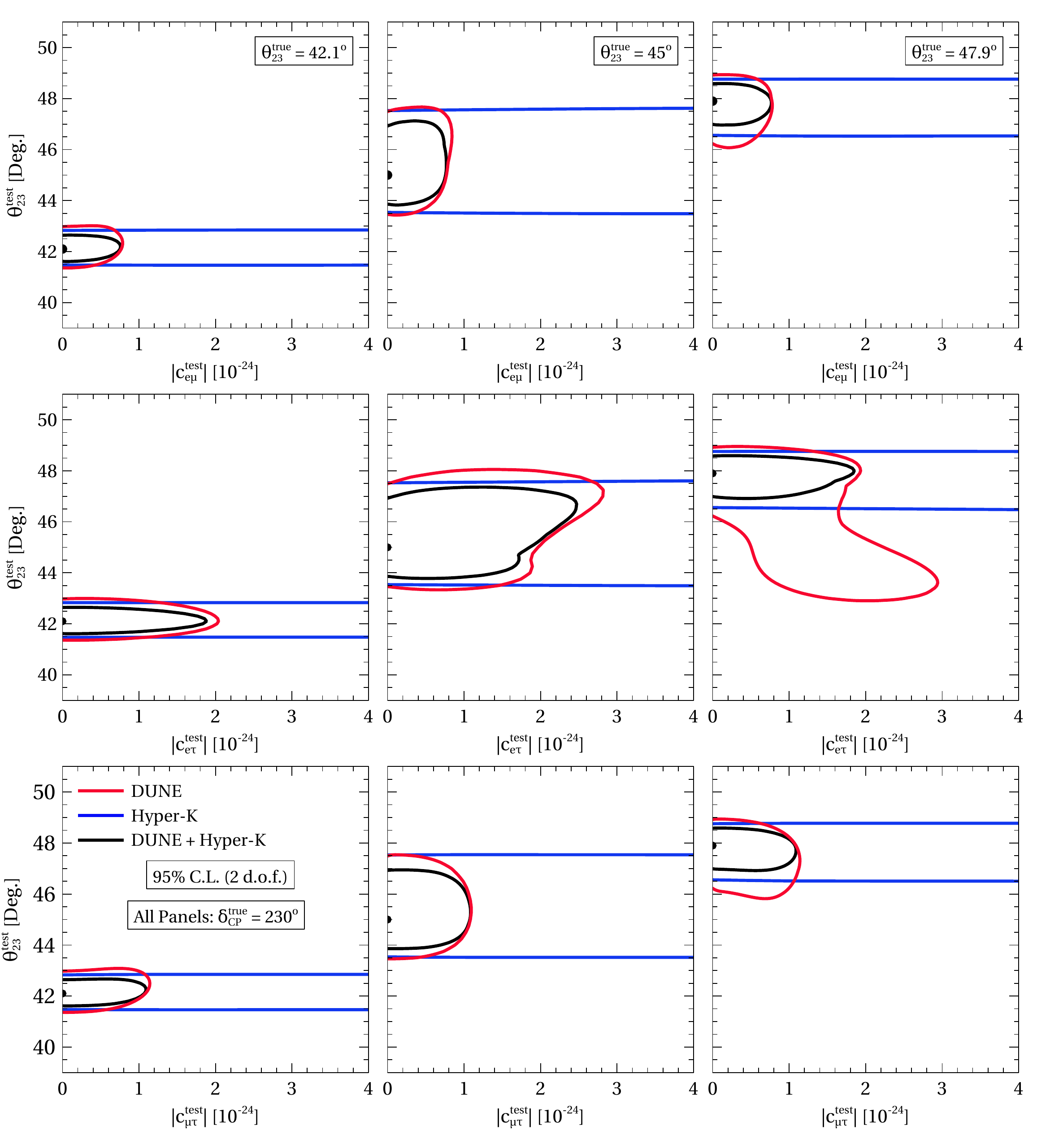}
	\vspace*{-10mm}
	\mycaption{95\% C.L. (2 d.o.f.) contours in the $\tzm-|c_{e\mu}|$ (top row), $\tzm-|c_{e\tau}|$ (middle row), and $\tzm-|c_{\mu\tau}|$ (bottom row) planes for DUNE, Hyper-K, and DUNE+Hyper-K. Three benchmark values of $\tzm$ considered in the data are $42.1^\circ$ (left column), $45^\circ$ (middle column), and $47.9^\circ$ (right column), as shown by black dots in each panel. True values of other oscillation parameters are given in Table~\ref{tab:params_value}. In the fit, we marginalize over $\delta_{\mathrm{CP}}$ in its allowed $3\sigma$ range and new phase $\phi$ in its entire allowed range.}
	\label{fig:correlation_th23_cptc}
\end{figure}

In Fig.~\ref{fig:correlation_dcp_cptv}, we show the correlations between the CPT-violating LIV parameters and the standard CP phase $\delta_{\rm CP}$. In the fit, we marginalize over $\theta_{23}$ in its allowed $3\sigma$ range~\cite{Esteban:2020cvm} and the phase $\phi$ associated with the off-diagonal LIV parameters in the range [0, $360^\circ$]. As mentioned earlier, all the other standard oscillation parameters are fixed at their best-fit values given in Table~\ref{tab:params_value}, both in data and theory. 
Top, middle, and bottom rows correspond to non-zero $a_{e\mu}$, $a_{e\tau}$, and $a_{\mu\tau}$, respectively, where we consider these LIV parameters 
one at-a-time in the fit. We take three different choices for the value of $\delta_{\rm CP}$ in the data that are allowed in the current $3\sigma$ limits, namely, $180^\circ$ (left panels), $270^\circ$ (middle panels), $300^\circ$ (right panels) as shown by the black dot in each panel. The red, blue, and black curves in each plot correspond to DUNE, Hyper-K, and the combination DUNE+Hyper-K, respectively. Each contour represents the allowed regions at 95\% C.L. (2 d.o.f.). We observe from the figure that for all the LIV parameters and all choices of the $\delta_{\mathrm{CP}}$ in the data, the allowed regions in $\delta_{\rm CP}-|a_{\alpha\beta}|$ planes are significantly small for DUNE compared to Hyper-K. One can understand it from the analytical expression of the oscillation probabilities discussed in  Sec.~\ref{sec:LIV}. From Eq.~\ref{eq:p_cptv_liv} and Eq.~\ref{eq:Pmm_liv}, we see that contribution from the CPT-violating LIV parameters is directly proportional to $L$. So, DUNE being an experiment with a comparatively longer baseline, shows better sensitivity to the CPT-violating LIV parameters. Hence, it has smaller allowed regions in  $\delta_{\rm CP}-|a_{\alpha\beta}|$ plane as compared to Hyper-K.

In case of $a_{e\mu}$, we notice a non-trivial degenerate solution in $\delta_{\rm CP}-|a_{e\mu}|$ plane which is centered around a non-zero value of  $|a_{e\mu}|$. This happens for both DUNE and Hyper-K at $|a_{e\mu}|\approx1\times 10^{-23}$ GeV and  $|a_{e\mu}|\approx4\times 10^{-23}$ GeV, respectively. This mainly occurs due to the degeneracy between $\theta_{23}$ and the complex phases ($\delta_{\rm CP}~\rm and~ \phi$) in the LIV contributing term in the appearance channel (see Eq.~\ref{eq:p_cptv_liv}), which plays a major role in constraining this parameter. For some combination of $\theta_{23}$ and $\delta_{\mathrm{CP}}+\phi$, this term minimizes at some non-zero value of $|a_{e\mu}|$ resulting in degeneracy with the standard oscillation case. As a result, we observe an allowed region at around that value of $|\aem|$. However, when we take the combined setup DUNE+Hyper-K, this degeneracy disappears. This happens because the values of $L$ and $E$ in the LIV contributing terms are now different for these two experiments, which helps in lifting the degeneracy. We also observe that upon combining these two setups, the allowed regions shrink further. In the case of $|\aet|$ (middle row) and $|\amt|$ (bottom row), we do not observe such degenerate solutions at 95\% C.L. (2 d.o.f.). In both cases, DUNE+Hyper-K shows a small improvement in the sensitivity as compared to DUNE. 

In Fig.~\ref{fig:correlation_dcp_cptc}, we repeat the above analysis for three off-diagonal CPT-conserving LIV parameters, $c_{e\mu}$ (top row), $c_{e\tau}$ (middle row), and $c_{\mu\tau}$ (bottom row). It is clear from the figure that DUNE shows a noticeable correlation between $c_{\alpha\beta}$ ($\alpha,\beta=e,\mu,\tau;\alpha\neq\beta$) and $\delta_{\mathrm{CP}}$. However, for Hyper-K, there is almost no correlation between those two parameters in all three cases. One can explain it from the fact that CPT-conserving parameters have negligible impact on both appearance and disappearance probabilities for Hyper-K as shown in the bottom row of Fig.~\ref{fig:cptc_app_prob} and bottom right panel of Fig.~\ref{fig:disapp_prob}. As discussed earlier,
this happens because of $L\times E$ dependencies in LIV contributing terms in the CPT-conserving case (see Eq.~\ref{eq:p_cptc_liv} and Eq.~\ref{eq:Pmm_liv}). Since Hyper-K has a shorter baseline and access to low energy neutrino beam compared to DUNE, it shows almost no sensitivity to CPT-conserving LIV parameters. 

In Fig.~\ref{fig:correlation_th23_cptv}, we show the correlation between the CPT-violating LIV parameters $a_{\alpha\beta}$ ($\alpha,\beta=e,\mu,\tau$; $\alpha\neq\beta$) and $\theta_{23}$. We consider three values of $\theta_{23}$ in data, namely, $42.1^\circ$ (left column) in the lower octant, 45$^\circ$ (middle column) maximal mixing case, and $47.9^\circ$ (right column) in the upper octant\footnote{We choose $\theta_{23}=42.1^\circ$ in lower octant as it is the current best-fit value from the global-fit of the oscillation parameters~\cite{Esteban:2020cvm}. For simulation, we marginalize over $\delta_{\mathrm{CP}}$ in its allowed $3\sigma$ range and the corresponding LIV phases in their entire allowed range. We consider the corresponding value ($47.9^\circ$) in the upper octant.}. We observe that for both DUNE and Hyper-K, the best result is obtained when the true value of $\theta_{23}$ is in the lower octant, where the allowed region is relatively small compared to the other two cases. 
Similar to $\delta_{\mathrm{CP}}-a_{\alpha\beta}$ correlation, DUNE performs significantly better compared to Hyper-K for all three choices of true $\theta_{23}$.
Here also, we observe degenerate allowed regions at non-zero values of $a_{e\mu}$ and $a_{e\tau}$ that appear at the opposite octant of $\theta_{23}$ for both DUNE and Hyper-K. This happens because of the degeneracy between $\theta_{23}$, $\delta_{\mathrm{CP}}$, and $\phi$ in the LIV contributing terms in the oscillation probabilities. On adding the data from DUNE and Hyper-K, the allowed regions become smaller, and interestingly, the degenerate regions appearing for the individual setups vanish, as shown by the black contour in each panel. In Fig.~\ref{fig:correlation_th23_cptc}, we show the same for CPT-conserving LIV parameters. Here also, the best results are obtained when $\theta_{23}$ is in lower octant for the individual setups. We observe that DUNE shows noticeable correlations in $\theta_{23}-|c_{\alpha\beta}|$ planes, but Hyper-K shows almost no correlation with the CPT-conserving LIV parameters. As mentioned before, Hyper-K has almost no sensitivity to the CPT-conserving LIV parameters because of its smaller baseline and lower neutrino energy.
However, we observe a slight improvement in the allowed regions when we combine the data from DUNE and Hyper-K.

\subsection{Constraints on CPT-violating and CPT-conserving LIV parameters}
\label{subsec:constraints}

In the previous section, we have discussed the correlations of the LIV parameters with the most uncertain standard oscillation parameters, $\delta_{\mathrm{CP}}$ and $\theta_{23}$ in the context of DUNE, Hyper-K, and their combination. In this section, we present the limits on the off-diagonal LIV parameters that would be obtained by these three setups. As discussed earlier, in our simulation, we marginalized over $\theta_{23}$, $\delta_{\text{CP}}$, and the phase associated with the off-diagonal LIV parameters in the fit (see Sec.~\ref{sec:RnA} for details). 

\begin{figure}[h!]
	\centering
	\includegraphics[width=\textwidth]{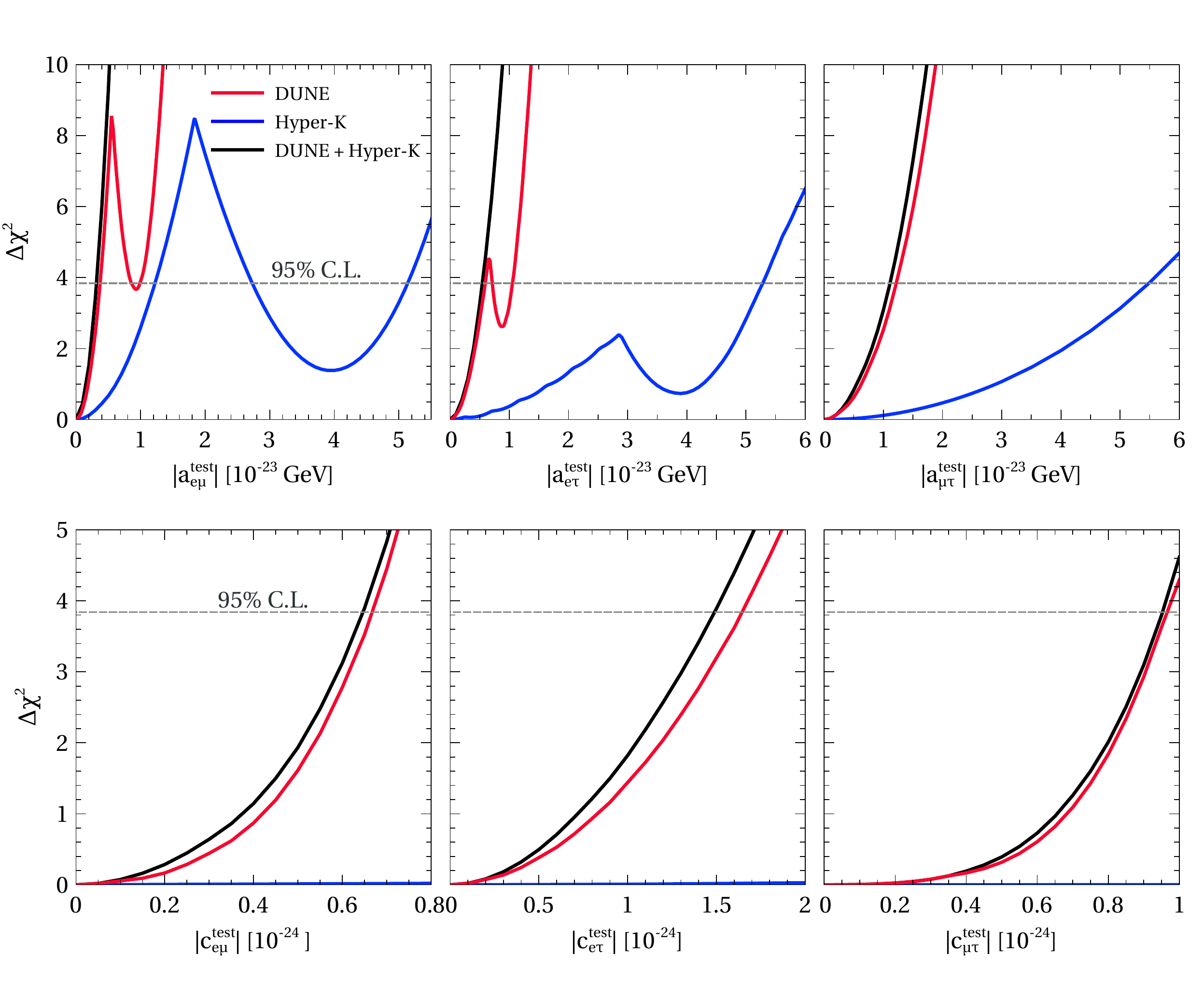}
	\vspace*{-10mm}
	\mycaption{ The expected limits on the CPT-violating (top row) and CPT-conserving (bottom row) LIV parameters
		from DUNE (red curves), Hyper-K (blue curves), and DUNE+Hyper-K (black curves).
		The upper (lower) panels show the $\Delta\chi^2$ for the off-diagonal CPT-violating (CPT-conserving) LIV parameters considering one-at-a-time. The true values of $\theta_{23}$ and $\delta_{\mathrm{CP}}$ are kept at their best fit values given in table~\ref{tab:params_value}. We marginalize over $\theta_{23}$ and $\delta_{\mathrm{CP}}$ in their 3$\sigma$ allowed range in the fit. Apart from $\theta_{23}$ and $\delta_{\mathrm{CP}}$, we also marginalize over the associated LIV phases in their total allowed range [$0^{\circ}$, $360^{\circ}$].}
	\label{fig:cptv_liv_bounds}
\end{figure}

In Fig.~\ref{fig:cptv_liv_bounds}, we show $\Delta \chi^2$ as a function of the off-diagonal CPT-violating (top row) and CPT-conserving (bottom row) LIV parameters. The red, blue, and black lines in each panel correspond to the sensitivity of DUNE, Hyper-K, and DUNE+Hyper-K setups, respectively. The top left panel correspond to $|a_{e\mu}|$, where we see that DUNE shows better sensitivity compared to Hyper-K at 95\% C.L.. Here, we find that for both DUNE and Hyper-K, there are local minima of $\Delta\chi^2$ around $1\times10^{-23}$ GeV and $4\times10^{-23}$ GeV, respectively. This feature can be explained using the correlations of the LIV parameters with standard oscillation parameters $\theta_{23}$ and $\delta_{\mathrm{CP}}$ discussed in Sec.~\ref{subsec:correlation}. We observe that there are degenerate allowed regions in $\delta_{\mathrm{CP}}-|a_{e\mu}|$ and $\theta_{23}-|a_{e\mu}|$ planes (see Figs.~\ref{fig:correlation_dcp_cptv} and \ref{fig:correlation_th23_cptv}) around the same values of $|\aem|$ ($\approx1\times10^{-23}$ GeV for DUNE and $\approx 4\times10^{-23}$ GeV for Hyper-K), where the local minima occur. Since this parameter is mainly constrained by the appearance channel, it hints towards a degeneracy between the appearance probability in absence of any new physics and the same in presence of the LIV for some combination of $\theta_{23}, \delta_{\mathrm{CP}}$ and new phase $\phi$. However, this degeneracy vanishes as we combine the data from DUNE and Hyper-K, giving a more stringent limit on $|\aem|$. In the top middle panel, we show the sensitivity for $|\aet|$. Here also, the local minima of $\Delta\chi^2$ are observed for the individual setup DUNE and Hyper-K, which again occur due to degeneracies between $\theta_{23}$, $\delta_{\mathrm{CP}}$, and $\phi$ (see top middle panel of Fig.~\ref{fig:correlation_dcp_cptv} and \ref{fig:correlation_th23_cptv}). Adding the data from the two experiments solves the issue of local minima.
Top right panel shows the constraints on $|\amt|$ for the three setups. We observe that DUNE gives significantly better limits for $|\amt|$ as compared to Hyper-K. Unlike $|\aem|$ and $|\aet|$, we do not observe any local minima of $\Delta\chi^2$ for $|\amt|$. It happens because $|\amt|$ is mainly constrained by the disappearance channel, where such degeneracy among $\theta_{23}$ and the CP phases ($\delta_{\rm CP}$ and $\phi$) does not occur.
In the lower panels of Fig.~\ref{fig:cptv_liv_bounds}, we show the constraints on CPT-conserving LIV parameters. As it is clear from the oscillation probability plots in Fig.~\ref{fig:cptc_app_prob} (see bottom row) and Fig.~\ref{fig:disapp_prob} (see bottom right panel), Hyper-K has almost no sensitivity on the CPT-conserving LIV parameters. However, when the data from Hyper-K and DUNE are added, sensitivities are slightly improved for all the three off-diagonal parameters.

In Table~\ref{tab:constraints_a}, we list the expected constraints on the off-diagonal CPT-violating and CPT-conserving LIV parameters at 95\% C.L. The second and third columns show the limits from DUNE and Hyper-K, respectively. The fourth column is the ultimate limit on LIV parameters from the combination of DUNE and Hyper-K.
Note that for the bounds on $|\aem|$ and $|\aet|$, we consider the most conservative scenarios, $\ie$ the largest value of $|a_{e\beta}|$ ($\beta = \mu,~\tau$), which reaches 95\% C.L. value. For $|a_{e\beta}|$, the obtained constraints from DUNE are almost five times better than that of Hyper-K. Also, combining the data from DUNE and Hyper-K, the limits improved further by a factor of $\approx 3$ for $|\aem|$  and $\approx 2$ for $|\aet|$ compared to DUNE. For $|\amt|$ also, constraints from DUNE outperform Hyper-K approximately by a factor of four. However, DUNE+Hyper-K setup improves the bounds on $|\amt|$ by only $\approx12\%$ compared to the standalone DUNE. In the case of CPT-conserving LIV parameters, the constraints from DUNE are incomparable to that of Hyper-K, as the latter shows almost no sensitivity to CPT-conserving LIV parameters. However, the combined setup DUNE+Hyper-K shows a marginal improvement in the limits compared to DUNE only.

\begin{table}[h!]
	\begin{center}
		\begin{adjustbox}{width=0.8\textwidth}
			\begin{tabular}{|c|c|c|c|c|}
				\hline\hline
				&DUNE& Hyper-K & DUNE+Hyper-K & T2K+NO$\nu$A\\ 
				\hline
				$|a_{e\mu}|~[10^{-23}~\rm {GeV}]$ &  $<$ 1.0 & $<$ 5.15 & $<$ 0.32 & $<$ 6.1\\ 
				\hline
				$|a_{e\tau}|~[10^{-23}~\rm {GeV}]$& $<$ 1.05 & $<$ 5.3 & $<$ 0.55& $<$ 7.0\\
				\hline
				$|a_{\mu\tau}|~[10^{-23}~\rm {GeV}]$& $<$ 1.26 & $<$ 5.5 & $<$ 1.1& $<$ 8.3\\
				\hline
				$|c_{e\mu}|~[10^{-24}]$ &  $<$ 0.66 &  $<$ 17.1  &   $<$ 0.64 & $<$ 11.0 \\ 
				\hline
				$|c_{e\tau}|~[10^{-24}]$& $<$ 1.65 &  $<$ 71.1 &   $<$ 1.49& $<$ 37.5\\
				\hline
				$|c_{\mu\tau}|~[10^{-24}]$& $<$ 0.97 &  $<$ 42.4 &   $<$ 0.95 & $<$ 29.0\\
				\hline\hline
			\end{tabular}
		\end{adjustbox}
		\mycaption{Expected bounds on the off-diagonal CPT-violating and CPT-conserving LIV parameters at 95\% C.L. (1 d.o.f.) using DUNE, Hyper-K, and the combination of DUNE and Hyper-K. Last column shows the results using the combination of T2K and NO$\nu$A with their full exposures.  }
		\label{tab:constraints_a}
	\end{center}
\end{table}

For a comparison with currently running long-baseline experiments T2K and NO$\nu$A, in the last column, we provide expected bounds from the combination of T2K and NO$\nu$A considering their full exposure. We assume a total exposure of 84.4 kt$\cdot$MW$\cdot$yrs for T2K~\cite{T2K:2001wmr, T2K:2011qtm, T2K:2014xyt} with five years of total runtime divided equally in neutrino and antineutrino modes. For NO$\nu$A~\cite{Ayres:2002ws, NOvA:2004blv, NOvA:2007rmc, Patterson:2012zs}, we consider total exposure of 58.8 kt$\cdot$MW$\cdot$yrs with six years of runtime with three years each in neutrino and antineutrino mode. We observe that DUNE puts significantly better constraints on both the CPT-violating and CPT-conserving LIV parameters than T2K+NO$\nu$A setup because the former has a larger baseline and better systematic uncertainties. However, the constraints on CPT-violating LIV parameters from T2K+NO$\nu$A setup are close to that of Hyper-K. Since NO$\nu$A has a comparatively larger baseline ($L=810$ km), it has the upper hand in putting stringent bounds on the CPT-violating parameters. However, Hyper-K has less systematic uncertainties to compensate for its small baseline. Similarly, for the CPT-conserving LIV parameters, the limits from T2K+NO$\nu$A setup are of the same order as Hyper-K, where the former gives slightly better constraints. It is because, apart from the larger baseline, the energy of the neutrino beam is also prominent for NO$\nu$A, which plays a vital role in constraining CPT-conserving LIV parameters.

The fourth column of Table~\ref{tab:constraints_a} shows the ultimate constraints on the off-diagonal CPT-violating and CPT-conserving LIV parameters that would be set from the combination of two next-generation long-baseline experiments DUNE and Hyper-K at 95\% C.L. In Table~\ref{tab:existing_bounds}, we show the existing bounds on some CPT-violating and CPT-conserving LIV parameters from the atmospheric neutrino  experiments, Super-K and IceCube. Comparing these with our results in Table~\ref{tab:constraints_a}, we observe that DUNE alone would be able to give better constraints of $\aem$ and $\aet$ as compared to the existing bounds from Super-K. Combining DUNE and Hyper-K can improve the constraints for $a_{e\beta}$ ($\beta=\mu,~\tau$) by almost one order of magnitude. One reason for this is the fact that $a_{e\beta}$ ($\beta=\mu,~\tau$) are mainly constrained by the appearance channel, which is the most important oscillation channel for a next-generation long-baseline experiment like DUNE that has a considerably larger baseline. For the atmospheric neutrino experiment like Super-K, the major channel is $\nu_\mu\to\nu_\mu$ disappearance channel, in which these two LIV parameters do not appear in the leading order. For $\amt$, projected results from the DUNE+Hyper-K setup are of the same order as Super-K, with the later having a slightly better limit. In the case of CPT-conserving LIV parameters, we observe that the existing limits from Super-K are at least one order better for $\cem$ and $\cmt$ compared to our results for DUNE+Hyper-K. However, for $\cet$, the expected limits from DUNE+Hyper-K are comparable with Super-K. This mainly happens because the contributions from CPT-conserving LIV parameters to oscillation probabilities in both the oscillation channels are proportional to both neutrino energy and the baseline. Atmospheric neutrino experiment like Super-K probes a significantly larger range of neutrino energies and baselines as compared to the LBL experiment like DUNE. So we expect Super-K to have better limits on the CPT-conserving LIV parameters.

\section{Summary and Conclusions}
\label{sec:SnC}

In the past few decades, data from outstanding neutrino oscillation experiments that are either completed or currently operational has almost settled the issue of measuring standard three-flavor neutrino oscillation parameters with excellent precision. Apart from resolving a few remaining issues in the three-neutrino paradigm, another major goal of the next-generation neutrino oscillation experiments will be to search for various physics beyond the standard model, which will open up a new era in particle physics. 
With that motivation, in this work, we probe the Lorentz invariance violation and its impact on neutrino flavor transition in the context of the two most anticipated upcoming long-baseline experiments, DUNE and Hyper-K. The Lorentz invariance violation can be realized in low energy effective field theories where the LIV interaction terms in the lagrangian come as multiplication of Lorentz violating coefficients and Lorentz violating operators of arbitrary mass dimensions. The coefficients of the dimension-three and dimension-four operators are, respectively, CPT-violating and CPT-conserving. In this work, for the first time, we have explored the CPT-conserving LIV parameters in the context of long-baseline experiments. Here, we focus on the isotropic components of the off-diagonal CPT-violating and CPT-conserving LIV parameters. The presence of non-zero CPT-violating and CPT-conserving parameters modify the neutrino propagation Hamiltonian and hence the oscillation probabilities, making them worth studying in neutrino oscillation experiments.

To have an analytical understanding about the impact of various LIV parameters on the neutrino oscillation probability, we use the perturbative approach to derive simple approximate analytical expressions for the $\nu_\mu\to\nu_e$ appearance and $\nu_\mu\to\nu_\mu$ disappearance probabilities. Here, we keep the terms up to first order in $\alpha$, $\sin^2\theta_{13}$, and LIV parameters $a_{\alpha\beta}/c_{\alpha\beta}$ ($\alpha,\beta = e,\mu,\tau;\alpha\neq\beta$). We find that for the appearance channel, $a_{e\mu}/c_{e\mu}$ and $a_{e\tau}/c_{e\tau}$ appear at the leading order, whereas for the disappearance channel, only $a_{\mu\tau}/c_{\mu\tau}$ presents. Our analytical expressions explain various features of oscillation probabilities shown in Figs.~[\ref{fig:app_prob}-\ref{fig:disapp_prob}], where we plot the exact oscillation probabilities numerically. We explain how the impact of LIV on oscillation probabilities depends on the values of phases associated with the off-diagonal LIV parameters using the LIV contributing terms in the oscillation probabilities given in Eqs.~\ref{eq:p_cptv_liv},~\ref{eq:p_cptc_liv}, and \ref{eq:Pmm_liv}. Also, we find that the LIV contributing terms in our analytical expressions for CPT-violating and CPT-conserving LIV parameters are proportional to $L$ and $L\times E$, respectively, both in the appearance and disappearance channels. As a result, DUNE, with a larger baseline and higher energy of the neutrino beam, shows significantly larger sensitivity to the LIV parameters than Hyper-K. As shown in lower panels of  Fig.~\ref{fig:cptc_app_prob}, Hyper-K shows negligible sensitivity to the CPT-conserving LIV parameters.

Using the configuration of the DUNE and Hyper-K as tabulated in Table~\ref{tab:exp_details}, we calculate the expected total event rates from the two setups in the standard case and in the presence of off-diagonal CPT-violating and CPT-conserving LIV parameters considered one-at-a-time (see Table~\ref{tab:total_events}). As expected from the probability analysis, we observe a major change in the event rate from the SI case in the presence of $a_{e\mu}$ ($a_{e\tau}$), which shows 41\% (49\%) deviation in the event rates for DUNE and 7.3\%(14\%) for Hyper-K when we consider $|a_{e\beta}|=2\times10^{-23}$ GeV ($\beta=\mu,\tau$). In the disappearance channel, the presence of $a_{\mu\tau}$ leads to modifications in the event rates by 7.8\% for DUNE and 2.2\% for Hyper-K. The CPT-conserving LIV parameters for which we consider the strength $|c_{\alpha\beta}|=1\times 10^{-24}$ show comparatively small changes in the event rates with a maximum 11\% changes for DUNE and $<1\%$ for Hyper-K.

We discuss the correlation between various LIV parameters and most uncertain oscillation parameters $\theta_{23}$ and $\delta_{\mathrm{CP}}$ (see Sec.~\ref{subsec:correlation}). To demonstrate this, we show allowed regions at 95\% C.L. (2 d.o.f.) in  $\delta_{\mathrm{CP}}-|a_{\alpha\beta}|/|c_{\alpha\beta}|$ (see Fig.~\ref{fig:correlation_dcp_cptv} and Fig.~\ref{fig:correlation_dcp_cptc})  and $\theta_{23}-|a_{\alpha\beta}|/|c_{\alpha\beta}|$ (see Fig.~\ref{fig:correlation_th23_cptv} and Fig.~\ref{fig:correlation_th23_cptc}) planes. In all cases, DUNE shows a more constrained allowed region in the above-mentioned planes, as expected from the probability plots. We notice almost no correlation between the CPT-conserving LIV parameters and standard oscillation parameters ($\theta_{23}$ and $\delta_{\rm CP}$) for Hyper-K. For $\delta_{\mathrm{CP}}-|a_{e\mu}|$ case, we observe some allowed regions centered around a non-zero value of $|a_{e\mu}|$, both for DUNE and Hyper-K. It happens due to degeneracy between the $\theta_{23}$ and the phases $\delta_{\mathrm{CP}}$ and $\phi$, which minimizes the contribution from LIV at some non-zero value of the $|a_{e\mu}|$. Also, in the case of $\theta_{23}-|a_{e\mu}|$ correlation, we observe such degenerate regions in opposite octant to the value of $\theta_{23}$ considered in the data. This also happens due to degeneracy between these three parameters ($\theta_{23}$, $\delta_{\mathrm{CP}}$ and $\phi$). In both the cases, the degenerate allowed regions vanish upon combining the data from DUNE and Hyper-K. Also, there is improvement in allowed regions in both $\delta_{\mathrm{CP}}-|a_{\alpha\beta}|$ and $\theta_{23}-|a_{\alpha\beta}|$ planes for the DUNE+Hyper-K setup. For the CPT-conserving case, this improvement is marginal. 
 
From the discussion on the correlation between the LIV parameters and standard oscillation parameters $\delta_{\mathrm{CP}}$ and $\theta_{23}$ in DUNE and Hyper-K, we get an idea about the sensitivity of these two setups to the off-diagonal CPT-violating and CPT-conserving LIV parameters. Here, we present the limits (see Fig.~\ref{fig:cptv_liv_bounds} and Table~\ref{tab:constraints_a}) on the LIV parameters that DUNE, Hyper-K, and their combination are expected to place with their full exposures. For CPT-violating LIV parameters, DUNE shows almost five times better constraints than Hyper-K.
Also, for $|a_{e\beta}|$ ($\beta = \mu,\tau$), we observe a local minimum in $\Delta\chi^2$ which results in deterioration in the constraints from DUNE and Hyper-K. It happens due to the degeneracy between $\theta_{23}$ and the phases $\delta_{\mathrm{CP}}$ and $\phi$ which is clear from the correlation plots. 
For the combination DUNE+Hyper-K, limits improve significantly for $|a_{e\beta}|$ ($\beta = \mu, \tau$) as the above discussed degeneracies vanish. For $a_{\mu\tau}$, DUNE outperforms Hyper-K, and their combination results in a small improvement in the limits. Hyper-K shows almost no sensitivity for the CPT-conserving LIV parameters. As a result, Hyper-K produces incomparably worse constraints relative to DUNE. To compare the limits from the future LBL experiments with the currently operating ones, we show the expected constraints by combining T2K and NO$\nu$A setups with their full exposures (see the last column of Table~\ref{tab:constraints_a}). We observe for CPT-violating LIV parameters, though the bounds from T2K+NO$\nu$A are worse than DUNE, it is comparable to that of Hyper-K. It is mainly because of the larger baseline of NO$\nu$A than Hyper-K that has better systematic uncertainties than the former. For the CPT-conserving parameters, T2K+NO$\nu$A gives slightly better constraints than Hyper-K, again because the former has a longer baseline and higher energy of the neutrino beam. We also compare our results with the existing limits on CPT-violating and CPT-conserving parameters listed in Table~\ref{tab:existing_bounds} from Super-K. We find that for CPT-violating parameters, especially $a_{e\beta}$ ($\beta=\mu,\tau$), projected limits at 95\% C.L. for the DUNE and Hyper-K combination is better compared to Super-K. It happens because the limits on $a_{e\beta}$ are mainly driven by the appearance channel, which is a major oscillation channel for LBL experiments, whereas the atmospheric neutrino experiment like Super-K mainly probe disappearance channel. For CPT-conserving parameters, Super-K shows almost two-order better constraints compared to the DUNE+Hyper-K setup. It is because of the large energy of the neutrinos in atmospheric neutrino experiments.

In this work, we discuss the impact of LIV on neutrino flavor transition probabilities in the context of next-generation long-baseline experiments DUNE and Hyper-K and explore the ability of these two setups to constrain the CPT-violating and CPT-conserving LIV parameters. We conclude that although these two setups have good complementarity in their configurations, a smaller baseline and lower energy of the neutrino beam for Hyper-K make it less favorable to probe LIV as compared to DUNE. However, combining DUNE and Hyper-K can improve the limits on LIV parameters up to a certain extent. We hope that our present study can be an important addition to the several interesting beyond the Standard Model scenarios which can be probed in the next-generation long-baseline neutrino oscillation experiments.


\subsubsection*{Acknowledgments}

We acknowledge the support from the Department of Atomic Energy (DAE), Govt. of India, under the Project Identification Number RIO 4001. S.K.A. is supported by the Young Scientist Research Grant [INSA/SP/YSP/144/2017/1578] from the Indian National Science Academy (INSA). S.K.A. acknowledges the financial support from the Swarnajayanti Fellowship (sanction order No. DST/SJF/PSA- 05/2019-20) provided by the Department of Science and Technology (DST), Govt. of India, and the Research Grant (sanction order No. SB/SJF/2020-21/21) provided by the Science and Engineering Research Board (SERB), Govt. of India, under the Swarnajayanti Fellowship project. We thank A. Raychaudhuri and M. Schreck for their insightful comments. We would also thank M. Singh and A. Kumar for useful communications. S.K.A would like to thank the United States-India Educational Foundation for providing the financial support through the Fulbright-Nehru Academic and Professional Excellence Fellowship (Award No. 2710/F-N APE/2021). The numerical simulations are carried out using SAMKHYA: High-Performance Computing Facility at Institute of Physics, Bhubaneswar.

\begin{appendix}
	\renewcommand\thefigure{A\arabic{figure}}
	\renewcommand\theHfigure{A\arabic{figure}}
	\renewcommand\theequation{A\arabic{equation}}
	\setcounter{figure}{0} 
	\setcounter{equation}{0}
\section{Comparison Between Numerical and Analytical Probabilities}
\label{appndx}

\begin{figure}[h!]
	\centering
	\includegraphics[width=0.86\textwidth]{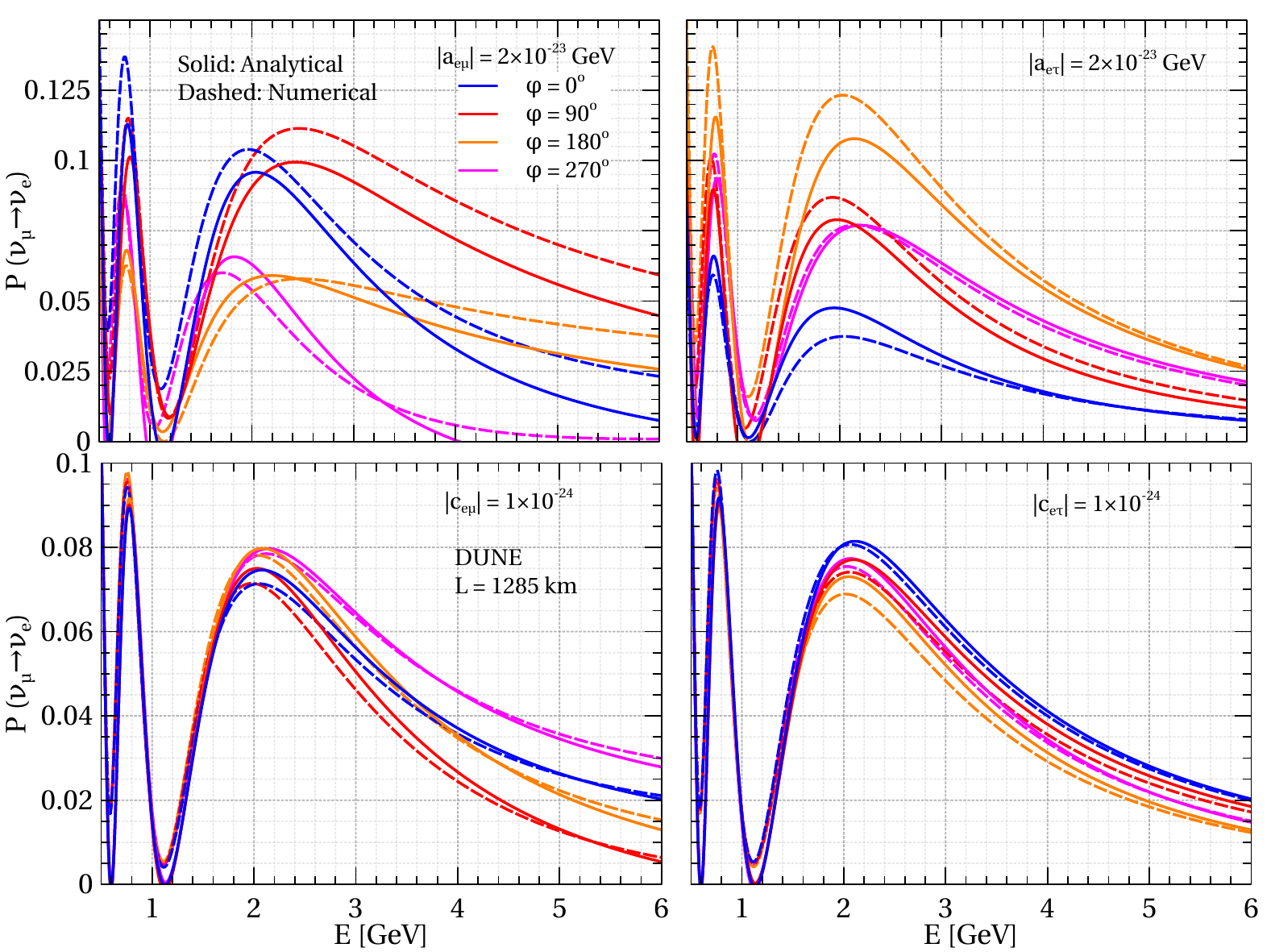}
	\mycaption{Comparison between exact $\nu_{\mu}\to\nu_e$ appearance probability calculated numerically (dashed lines) with the same calculated analytically (solid lines) using Eq.~\ref{eq:pme_liv} for a baseline $L=1285$ km. The top (bottom) row corresponds to the probability in the presence of CPT-violating (CPT-conserving) LIV parameters $a_{e\beta}$ ($c_{e\beta}$) with $\beta=\mu $ in left column, and $\beta=\tau$ in right column. Four colored curves correspond to four values of the associated phase, as mentioned in the legend. The values of the standard oscillation parameters used in the plot are given in Table~\ref{tab:params_value} with NMO.\label{fig:Pme-comparison}}
\end{figure}

In this section, we check the accuracy of the approximate analytical expressions of the oscillation probabilities derived in Sec.~\ref{sec:LIV}. In Fig.~\ref{fig:Pme-comparison}, we compare the $\nu_\mu\rightarrow\nu_e$ appearance probability in the presence of $a_{e\mu}/c_{e\mu}$ and $a_{e\tau}/c_{e\tau}$, calculated using the analytical expression with the exact oscillation probability calculated numerically using GLoBES software, for DUNE ($L = 1285$ km). We consider only these LIV parameters since they appear in the first order in the expansion parameters. We show the results for four values of the phase associated with the LIV parameters, namely, $0^\circ$, $90^\circ$, $180^\circ$, and $270^\circ$. We observe that for the CPT-violating LIV parameters, matching with the exact appearance probability is slightly worse compared to the CPT-conserving case, as the assumed strength of LIV parameters in the CPT-violating case is one order larger than the corresponding CPT-conserving case.

\begin{figure}[h!]
	\centering
	\includegraphics[width=0.86\textwidth]{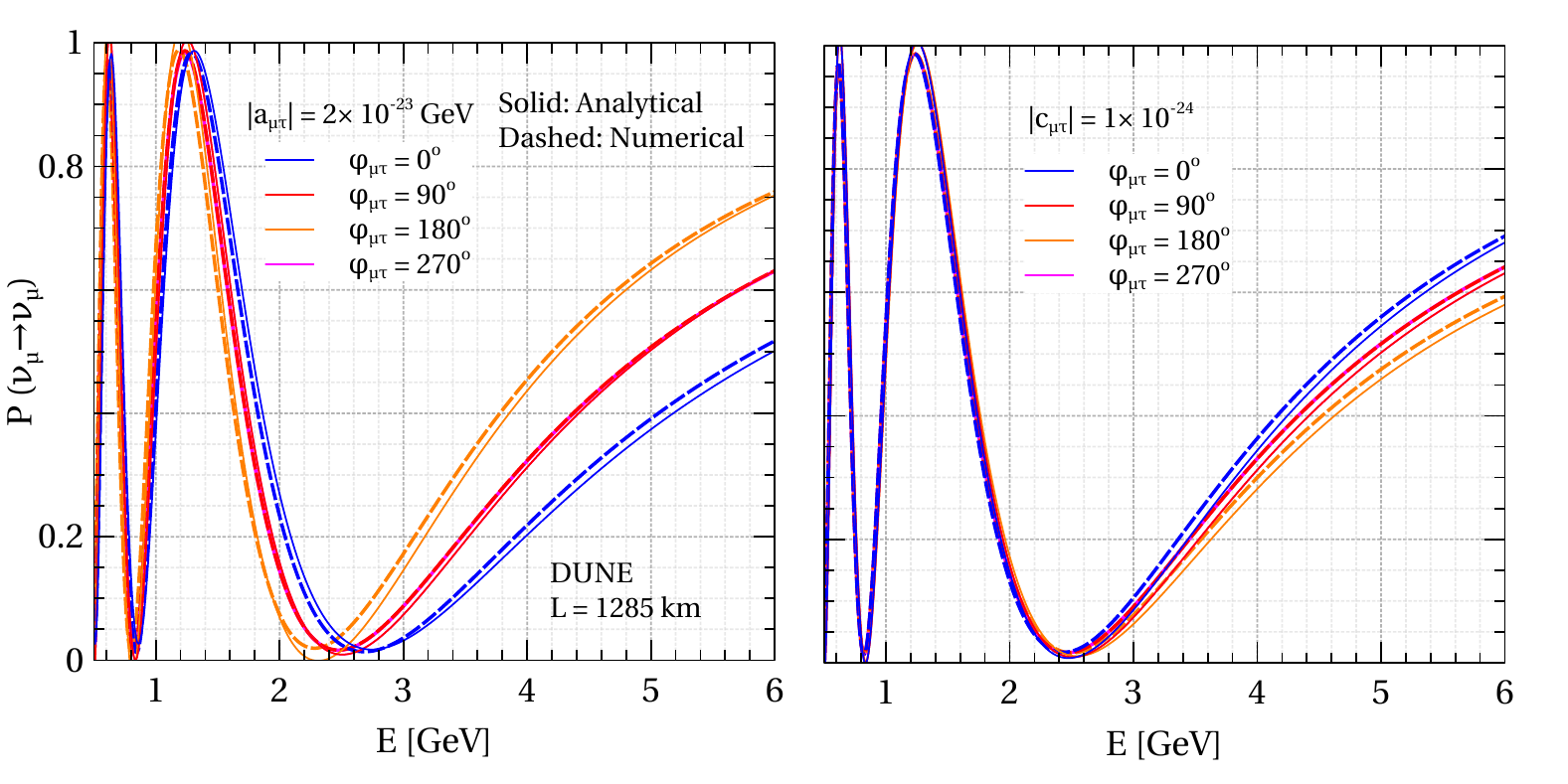}
	\mycaption{Comparison between exact $\nu_\mu\to\nu_{\mu}$ disappearance probability calculated numerically (dashed lines) with the same calculated analytically (solid lines) using Eq.~\ref{eq:pmm} for a baseline $L=1285$ km. The Left (right) column corresponds to the probability in the presence of CPT-violating (CPT-conserving) LIV parameter $a_{\mu\tau}$ ($c_{\mu\tau}$). Four colored curves correspond to four values of the associated phase, as mentioned in the legend. The values of the standard oscillation parameters used in the plot are given in Table~\ref{tab:params_value} with NMO.\label{fig:Pmm-comparison}}
\end{figure}

However, the features of the appearance probabilities for different values of LIV-phases are preserved by the analytical expression. In Fig.~\ref{fig:Pmm-comparison}, we show the same for the $\nu_\mu\to\nu_\mu$ disappearance probability. Here, we show the effect from the LIV parameters, $a_{\mu\tau}/c_{\mu\tau}$, since only these terms appear till the first order. Again, we consider four values of associated phases as in Fig.~\ref{fig:Pme-comparison}. We find the oscillation probabilities calculated analytically match quite well with the numerical ones. It is valid for all assumed values of the associated phases, both in CPT-violating and CPT-conserving scenarios. The same is true for Hyper-K.

\renewcommand\thefigure{B\arabic{figure}}
\renewcommand\theHfigure{B\arabic{figure}}
\renewcommand\theequation{B\arabic{equation}}
\setcounter{figure}{0} 
\setcounter{equation}{0}
\section{$\nu_{e}$ appearance and $\nu_{\mu}$ disappearance event spectra in DUNE and Hyper-K in presence of the CPT-violating LIV parameters}
\label{appndx-B}

\begin{figure}[h!]
	\centering
	\includegraphics[width=\textwidth]{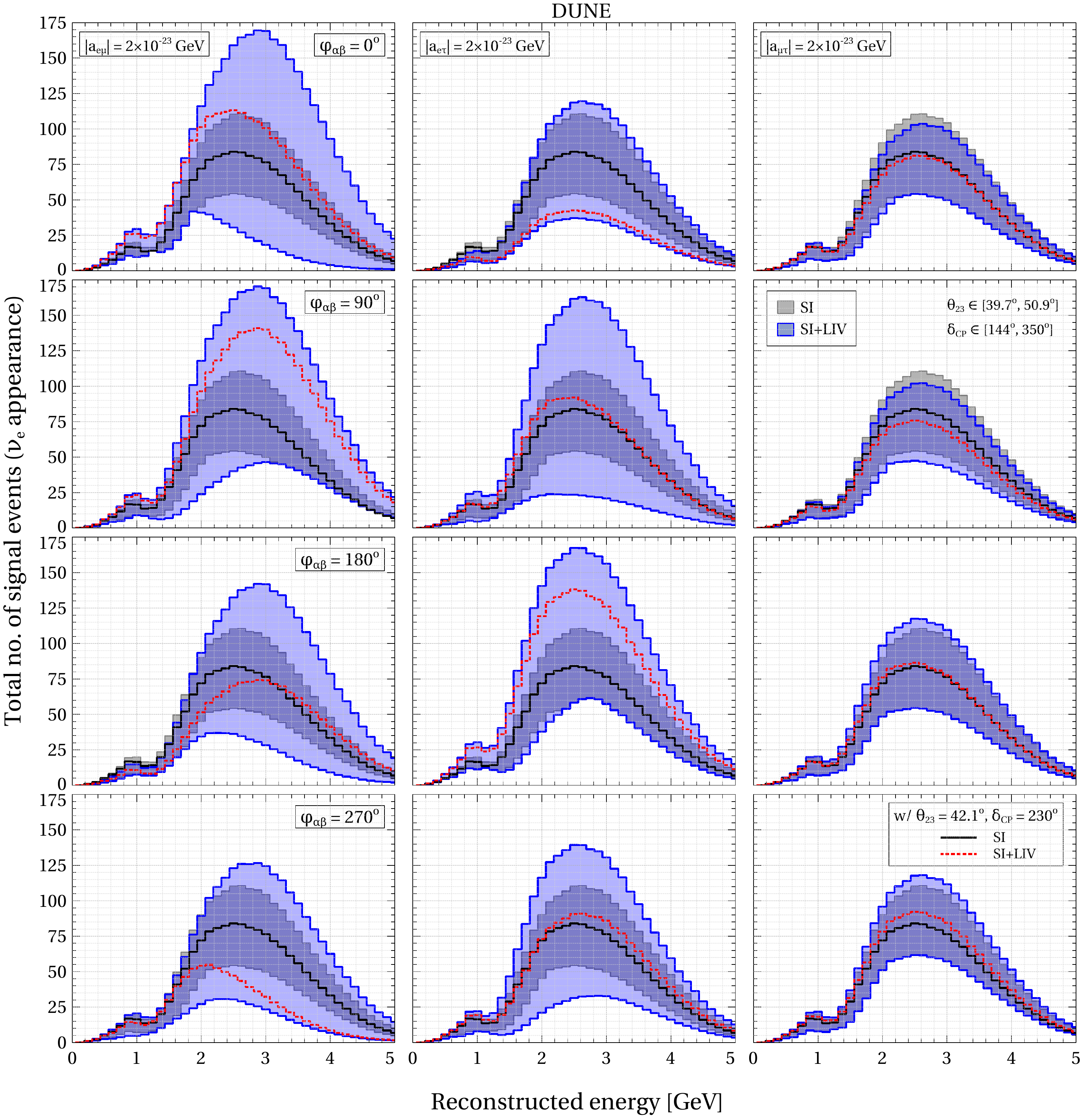}
	\vspace*{-5mm}
	\mycaption{The distribution band of reconstructed $\nu_e$ events (signal) obtained via $\nu_\mu \to \nu_e$ channel for both SI and SI+LIV scenarios at DUNE. The bands in each panel appear due to the uncertainties in the standard oscillation parameters, $\theta_{23}$ and $\delta_{\rm CP}$ in their $3\sigma$ range. The gray bands correspond to SI scenarios, whereas the blue ones represent CPT-violating LIV scenarios. The solid black and the dashed red lines in each panel represent the event spectra in case of SI and SI+LIV, respectively, with the oscillation parameters fixed at their benchmark values given in Table~\ref{tab:params_value}. The strength of these LIV parameters: $|a_{e\mu}|$ (first column), $|a_{e\tau}|$ (second column), and $|a_{\mu\tau}|$ (third column) are considered to be $2\times 10^{-23}$ GeV for four choices of the corresponding LIV phases, namely, $\phi_{\alpha\beta}$ = $0^\circ$ (first row), $90^\circ$ (second row), $180^\circ$ (third row), and $270^\circ$ (fourth row).\label{fig:event_rate_nu_app_dune}}
\end{figure}

\begin{figure}[h!]
	\centering
	\includegraphics[width=\textwidth]{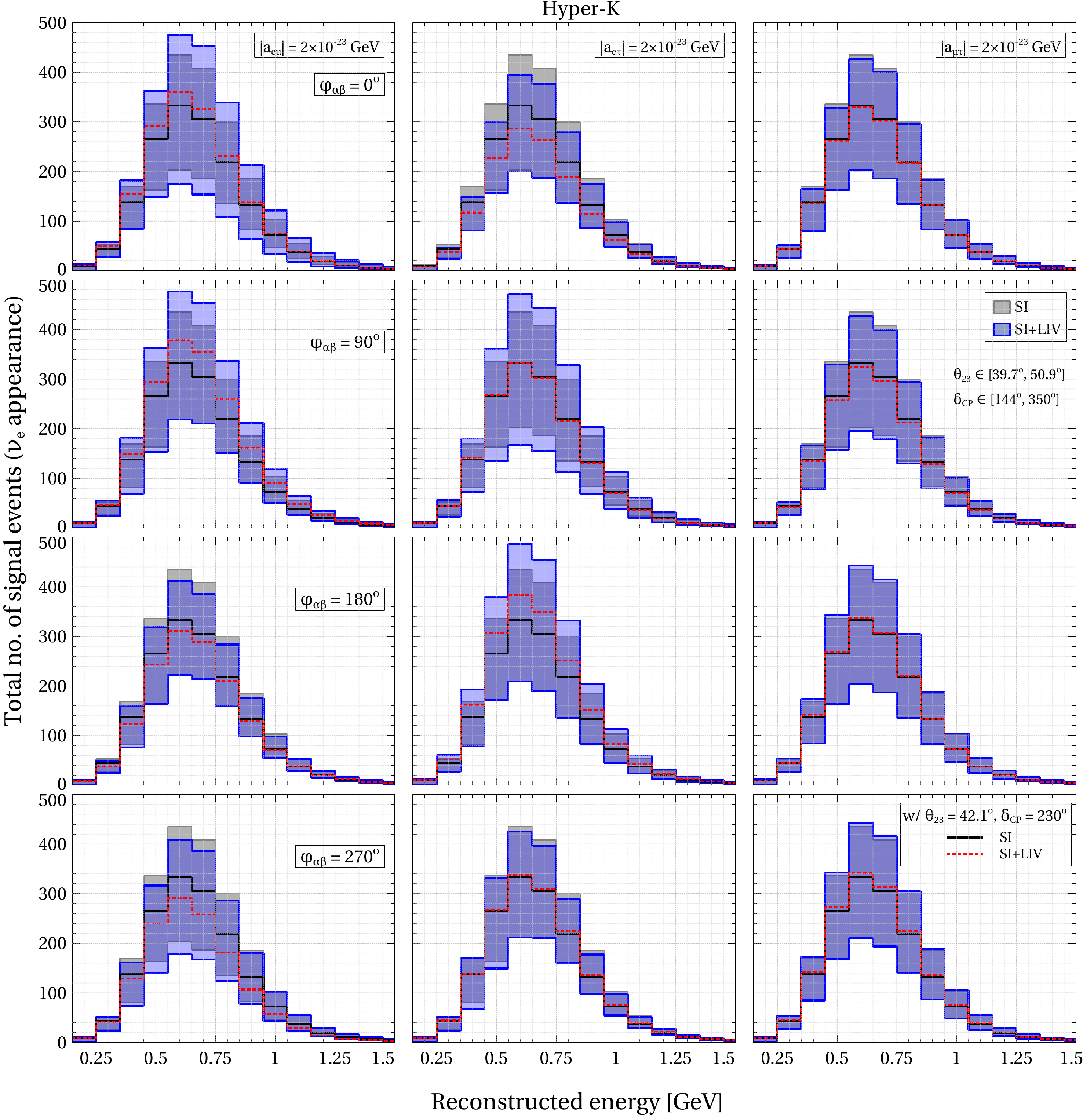}
	\vspace*{-5mm}
\mycaption{The distribution band of reconstructed $\nu_e$ events (signal) obtained via $\nu_\mu \to \nu_e$ channel for both SI and SI+LIV scenarios at Hyper-K. The bands in each panel appear due to the uncertainties in the standard oscillation parameters, $\theta_{23}$ and $\delta_{\rm CP}$ in their $3\sigma$ range. The gray bands correspond to SI scenarios, whereas the blue ones represent CPT-violating LIV scenarios. The solid black and the dashed red lines in each panel represent the event spectra in case of SI and SI+LIV, respectively, with the oscillation parameters fixed at their benchmark values given in Table~\ref{tab:params_value}. The strength of these LIV parameters: $|a_{e\mu}|$ (first column), $|a_{e\tau}|$ (second column), and $|a_{\mu\tau}|$ (third column) are considered to be $2\times 10^{-23}$ GeV for four choices of the corresponding LIV phases, namely, $\phi_{\alpha\beta}$ = $0^\circ$ (first row), $90^\circ$ (second row), $180^\circ$ (third row), and $270^\circ$ (fourth row). \label{fig:event_rate_nu_app_hk}}	
\end{figure}

\begin{figure}[h!]
	\centering
	\includegraphics[width=0.9\textwidth]{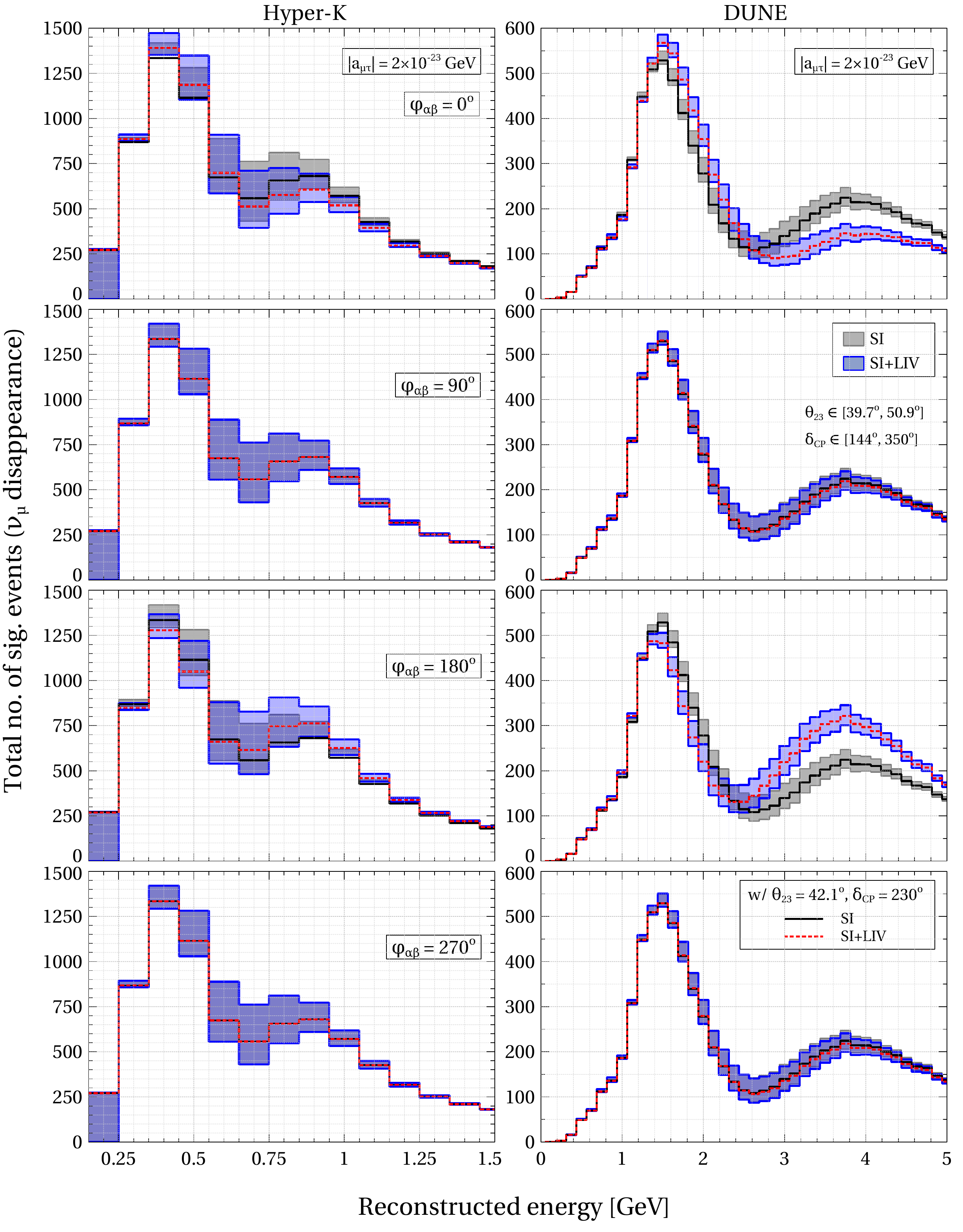}
	\vspace*{-2.5mm}
	\mycaption{The distribution band of reconstructed $\nu_\mu$ events (signal) obtained via $\nu_\mu \to \nu_\mu$ channel for both SI and SI+LIV scenarios at Hyper-K (left column) and DUNE (right column). The bands in each panel appear due to the uncertainties in the standard oscillation parameters, $\theta_{23}$ and $\delta_{\rm CP}$ in their $3\sigma$ range. The gray bands correspond to SI scenarios, whereas the blue ones represent CPT-violating LIV scenarios. The solid black and the dashed red lines in each panel represent the event spectra in case of SI and SI+LIV, respectively, with the oscillation parameters fixed at their benchmark values given in Table~\ref{tab:params_value}. Here, we consider only the effect of non-zero $|a_{\mu\tau}|$ whose value is considered to be $2\times 10^{-23}$ GeV for four choices of the corresponding LIV phase, namely, $\phi_{\mu\tau}$ = $0^\circ$ (first row), $90^\circ$ (second row), $180^\circ$ (third row), and $270^\circ$ (fourth row).\label{fig:event_rate_nu_disapp_hk_dune}}	
\end{figure}

In Figs.~\ref{fig:event_rate_nu_app_dune} and \ref{fig:event_rate_nu_app_hk}, we show the $\nu_e$ appearance signal event spectra in DUNE and Hyper-K, respectively, in presence of the CPT-violating LIV parameters, $a_{e\mu}$, $a_{e\tau}$, and $a_{\mu\tau}$. Since $a_{\mu\tau}$ mainly affects the $\nu_\mu\to\nu_{\mu}$ disappearance channel, we show the disappearance event spectra in Fig.~\ref{fig:event_rate_nu_disapp_hk_dune}, for Hyper-K and DUNE for non-zero $a_{\mu\tau}$. The gray band in each panel shows the neutrino signal rate considering $3\sigma$ uncertainties in the present best-fit values of the standard oscillation parameters~\cite{Esteban:2020cvm}, $\theta_{23}$ and $\delta_{\rm CP}$ in the SI case. Blue bands show the same in the presence of the off-diagonal CPT-violating LIV parameters one-at-a-time for four choices of the associated phase, namely, $0^\circ$, $90^\circ$, $180^\circ$, and $270^\circ$. The benchmark values of the LIV parameters are mentioned in the top panels. Features in the signal rate bands reflect the features in the oscillation probabilities shown in Figs.~\ref{fig:app_prob} and \ref{fig:disapp_prob}. 

In Fig.~\ref{fig:event_rate_nu_app_dune}, we observe that in the presence of $|a_{e\mu}|$ and $|a_{e\tau}|$,  the $\nu_e$ appearance signal rates exceed the SI band with present $3\sigma$ uncertainties in the oscillation parameters. This is valid for all four values of the associated phases. For $|a_{\mu\tau}|$, at the mentioned strength, deviation is small, as this parameter appears in the higher order terms in the appearance probability expression. For Hyper-K (see Fig.~\ref{fig:event_rate_nu_app_hk}), all deviations from expected signal rates in the SI case are comparatively smaller for all choices of the associated phase. The deviation is almost negligible for $|a_{\mu\tau}|$. It is expected from the fact that Hyper-K operates at lower energy and baseline as compared to DUNE. 

Similarly, from Fig.~\ref{fig:event_rate_nu_disapp_hk_dune}, it is evident that the deviation between the $\nu_\mu$ disappearance events rates, in SI case and in the presence of $a_{\mu\tau}$, is significant for DUNE. 
Hence, it is clear from these three figures that DUNE will have comparatively better potential to probe off-diagonal LIV parameters due to its access to higher neutrino energy and longer baseline.

\end{appendix}

\bibliographystyle{JHEP}
\bibliography{LBL_LV.bib}

\providecommand{\href}[2]{#2}\begingroup\raggedright\begin{thebibliography}{100}

\bibitem{Workman:2022ynf}
{\bf Particle Data Group} Collaboration, R.~L. Workman and Others, {\it {Review
  of Particle Physics}},  {\em PTEP} {\bf 2022} (2022) 083C01.

\bibitem{Super-Kamiokande:2004orf}
{\bf Super-Kamiokande} Collaboration, Y.~Ashie et~al., {\it {Evidence for an
  oscillatory signature in atmospheric neutrino oscillation}},  {\em Phys. Rev.
  Lett.} {\bf 93} (2004) 101801,
  [\href{http://arxiv.org/abs/hep-ex/0404034}{{\tt hep-ex/0404034}}].

\bibitem{Mohapatra:2005wg}
R.~N. Mohapatra et~al., {\it {Theory of neutrinos: A White paper}},  {\em Rept.
  Prog. Phys.} {\bf 70} (2007) 1757--1867,
  [\href{http://arxiv.org/abs/hep-ph/0510213}{{\tt hep-ph/0510213}}].

\bibitem{Strumia:2006db}
A.~Strumia and F.~Vissani, {\it {Neutrino masses and mixings and...}},
  \href{http://arxiv.org/abs/hep-ph/0606054}{{\tt hep-ph/0606054}}.

\bibitem{Gonzalez-Garcia:2007dlo}
M.~C. Gonzalez-Garcia and M.~Maltoni, {\it {Phenomenology with Massive
  Neutrinos}},  {\em Phys. Rept.} {\bf 460} (2008) 1--129,
  [\href{http://arxiv.org/abs/0704.1800}{{\tt arXiv:0704.1800}}].

\bibitem{Fantini:2018itu}
G.~Fantini, A.~Gallo~Rosso, F.~Vissani, and V.~Zema, {\it {Introduction to the
  Formalism of Neutrino Oscillations}},  {\em Adv. Ser. Direct. High Energy
  Phys.} {\bf 28} (2018) 37--119, [\href{http://arxiv.org/abs/1802.05781}{{\tt
  arXiv:1802.05781}}].

\bibitem{Super-Kamiokande:2016yck}
{\bf Super-Kamiokande} Collaboration, K.~Abe et~al., {\it {Solar Neutrino
  Measurements in Super-Kamiokande-IV}},  {\em Phys. Rev. D} {\bf 94} (2016),
  no.~5 052010, [\href{http://arxiv.org/abs/1606.07538}{{\tt
  arXiv:1606.07538}}].

\bibitem{SNO:2011hxd}
{\bf SNO} Collaboration, B.~Aharmim et~al., {\it {Combined Analysis of all
  Three Phases of Solar Neutrino Data from the Sudbury Neutrino Observatory}},
  {\em Phys. Rev. C} {\bf 88} (2013) 025501,
  [\href{http://arxiv.org/abs/1109.0763}{{\tt arXiv:1109.0763}}].

\bibitem{BOREXINO:2014pcl}
{\bf BOREXINO} Collaboration, G.~Bellini et~al., {\it {Neutrinos from the
  primary proton\textendash{}proton fusion process in the Sun}},  {\em Nature}
  {\bf 512} (2014), no.~7515 383--386.

\bibitem{Super-Kamiokande:2010tar}
{\bf Super-Kamiokande} Collaboration, K.~Abe et~al., {\it {Solar neutrino
  results in Super-Kamiokande-III}},  {\em Phys. Rev. D} {\bf 83} (2011)
  052010, [\href{http://arxiv.org/abs/1010.0118}{{\tt arXiv:1010.0118}}].

\bibitem{Super-Kamiokande:2017yvm}
{\bf Super-Kamiokande} Collaboration, K.~Abe et~al., {\it {Atmospheric neutrino
  oscillation analysis with external constraints in Super-Kamiokande I-IV}},
  {\em Phys. Rev. D} {\bf 97} (2018), no.~7 072001,
  [\href{http://arxiv.org/abs/1710.09126}{{\tt arXiv:1710.09126}}].

\bibitem{Super-Kamiokande:2019gzr}
{\bf Super-Kamiokande} Collaboration, M.~Jiang et~al., {\it {Atmospheric
  Neutrino Oscillation Analysis with Improved Event Reconstruction in
  Super-Kamiokande IV}},  {\em PTEP} {\bf 2019} (2019), no.~5 053F01,
  [\href{http://arxiv.org/abs/1901.03230}{{\tt arXiv:1901.03230}}].

\bibitem{IceCube:2017lak}
{\bf IceCube} Collaboration, M.~G. Aartsen et~al., {\it {Measurement of
  Atmospheric Neutrino Oscillations at 6\textendash{}56 GeV with IceCube
  DeepCore}},  {\em Phys. Rev. Lett.} {\bf 120} (2018), no.~7 071801,
  [\href{http://arxiv.org/abs/1707.07081}{{\tt arXiv:1707.07081}}].

\bibitem{ANTARES:2018rtf}
{\bf ANTARES} Collaboration, A.~Albert et~al., {\it {Measuring the atmospheric
  neutrino oscillation parameters and constraining the 3+1 neutrino model with
  ten years of ANTARES data}},  {\em JHEP} {\bf 06} (2019) 113,
  [\href{http://arxiv.org/abs/1812.08650}{{\tt arXiv:1812.08650}}].

\bibitem{DayaBay:2018yms}
{\bf Daya Bay} Collaboration, D.~Adey et~al., {\it {Measurement of the Electron
  Antineutrino Oscillation with 1958 Days of Operation at Daya Bay}},  {\em
  Phys. Rev. Lett.} {\bf 121} (2018), no.~24 241805,
  [\href{http://arxiv.org/abs/1809.02261}{{\tt arXiv:1809.02261}}].

\bibitem{RENO:2018dro}
{\bf RENO} Collaboration, G.~Bak et~al., {\it {Measurement of Reactor
  Antineutrino Oscillation Amplitude and Frequency at RENO}},  {\em Phys. Rev.
  Lett.} {\bf 121} (2018), no.~20 201801,
  [\href{http://arxiv.org/abs/1806.00248}{{\tt arXiv:1806.00248}}].

\bibitem{DoubleChooz:2019qbj}
{\bf Double Chooz} Collaboration, H.~de~Kerret et~al., {\it {Double Chooz
  $\theta_{13}$ measurement via total neutron capture detection}},  {\em Nature
  Phys.} {\bf 16} (2020), no.~5 558--564,
  [\href{http://arxiv.org/abs/1901.09445}{{\tt arXiv:1901.09445}}].

\bibitem{MINOS:2013utc}
{\bf MINOS} Collaboration, P.~Adamson et~al., {\it {Measurement of Neutrino and
  Antineutrino Oscillations Using Beam and Atmospheric Data in MINOS}},  {\em
  Phys. Rev. Lett.} {\bf 110} (2013), no.~25 251801,
  [\href{http://arxiv.org/abs/1304.6335}{{\tt arXiv:1304.6335}}].

\bibitem{T2K:2019bcf}
{\bf T2K} Collaboration, K.~Abe et~al., {\it {Constraint on the
  matter\textendash{}antimatter symmetry-violating phase in neutrino
  oscillations}},  {\em Nature} {\bf 580} (2020), no.~7803 339--344,
  [\href{http://arxiv.org/abs/1910.03887}{{\tt arXiv:1910.03887}}]. [Erratum:
  Nature 583, E16 (2020)].

\bibitem{NOvA:2019cyt}
{\bf NOvA} Collaboration, M.~A. Acero et~al., {\it {First Measurement of
  Neutrino Oscillation Parameters using Neutrinos and Antineutrinos by NOvA}},
  {\em Phys. Rev. Lett.} {\bf 123} (2019), no.~15 151803,
  [\href{http://arxiv.org/abs/1906.04907}{{\tt arXiv:1906.04907}}].

\bibitem{NOvA:2021nfi}
{\bf NOvA} Collaboration, M.~A. Acero et~al., {\it {An Improved Measurement of
  Neutrino Oscillation Parameters by the NOvA Experiment}},
  \href{http://arxiv.org/abs/2108.08219}{{\tt arXiv:2108.08219}}.

\bibitem{DUNE:2015lol}
{\bf DUNE} Collaboration, R.~Acciarri et~al., {\it {Long-Baseline Neutrino
  Facility (LBNF) and Deep Underground Neutrino Experiment (DUNE)}: {Conceptual
  Design Report, Volume 2: The Physics Program for DUNE at LBNF}},
  \href{http://arxiv.org/abs/1512.06148}{{\tt arXiv:1512.06148}}.

\bibitem{DUNE:2020lwj}
{\bf DUNE} Collaboration, B.~Abi et~al., {\it {Deep Underground Neutrino
  Experiment (DUNE), Far Detector Technical Design Report, Volume I
  Introduction to DUNE}},  {\em JINST} {\bf 15} (2020), no.~08 T08008,
  [\href{http://arxiv.org/abs/2002.02967}{{\tt arXiv:2002.02967}}].

\bibitem{DUNE:2020ypp}
{\bf DUNE} Collaboration, B.~Abi et~al., {\it {Deep Underground Neutrino
  Experiment (DUNE), Far Detector Technical Design Report, Volume II: DUNE
  Physics}},  \href{http://arxiv.org/abs/2002.03005}{{\tt arXiv:2002.03005}}.

\bibitem{DUNE:2020jqi}
{\bf DUNE} Collaboration, B.~Abi et~al., {\it {Long-baseline neutrino
  oscillation physics potential of the DUNE experiment}},  {\em Eur. Phys. J.
  C} {\bf 80} (2020), no.~10 978, [\href{http://arxiv.org/abs/2006.16043}{{\tt
  arXiv:2006.16043}}].

\bibitem{DUNE:2021cuw}
{\bf DUNE} Collaboration, B.~Abi et~al., {\it {Experiment Simulation
  Configurations Approximating DUNE TDR}},
  \href{http://arxiv.org/abs/2103.04797}{{\tt arXiv:2103.04797}}.

\bibitem{DUNE:2021mtg}
{\bf DUNE} Collaboration, A.~A. Abud et~al., {\it {Low exposure long-baseline
  neutrino oscillation sensitivity of the DUNE experiment}},
  \href{http://arxiv.org/abs/2109.01304}{{\tt arXiv:2109.01304}}.

\bibitem{Agarwalla:2021bzs}
S.~K. Agarwalla, R.~Kundu, S.~Prakash, and M.~Singh, {\it {A close look on 2-3
  mixing angle with DUNE in light of current neutrino oscillation data}},  {\em
  JHEP} {\bf 03} (2022) 206, [\href{http://arxiv.org/abs/2111.11748}{{\tt
  arXiv:2111.11748}}].

\bibitem{Agarwalla:2022xdo}
S.~K. Agarwalla, S.~Das, A.~Giarnetti, D.~Meloni, and M.~Singh, {\it {Enhancing
  Sensitivity to Leptonic CP Violation using Complementarity among DUNE, T2HK,
  and T2HKK}},  \href{http://arxiv.org/abs/2211.10620}{{\tt arXiv:2211.10620}}.

\bibitem{Hyper-KamiokandeWorkingGroup:2014czz}
{\bf Hyper-Kamiokande Working Group} Collaboration, K.~Abe et~al., {\it {A Long
  Baseline Neutrino Oscillation Experiment Using J-PARC Neutrino Beam and
  Hyper-Kamiokande}},  in {\em {18th J-PARC PAC meeting in May 2014}}, 12,
  2014.
\newblock \href{http://arxiv.org/abs/1412.4673}{{\tt arXiv:1412.4673}}.

\bibitem{Abe:2015zbg}
{\bf Hyper-Kamiokande Proto-Collaboration} Collaboration, K.~Abe et~al., {\it
  {Physics potential of a long-baseline neutrino oscillation experiment using a
  J-PARC neutrino beam and Hyper-Kamiokande}},  {\em PTEP} {\bf 2015} (2015)
  053C02, [\href{http://arxiv.org/abs/1502.05199}{{\tt arXiv:1502.05199}}].

\bibitem{Hyper-Kamiokande:2018ofw}
{\bf Hyper-Kamiokande} Collaboration, K.~Abe et~al., {\it {Hyper-Kamiokande
  Design Report}},  \href{http://arxiv.org/abs/1805.04163}{{\tt
  arXiv:1805.04163}}.

\bibitem{Arguelles:2019xgp}
C.~A. Arg\"uelles et~al., {\it {New opportunities at the next-generation
  neutrino experiments I: BSM neutrino physics and dark matter}},  {\em Rept.
  Prog. Phys.} {\bf 83} (2020), no.~12 124201,
  [\href{http://arxiv.org/abs/1907.08311}{{\tt arXiv:1907.08311}}].

\bibitem{Arguelles:2022tki}
C.~A. Arg\"uelles et~al., {\it {Snowmass white paper: beyond the standard model
  effects on neutrino flavor: Submitted to the proceedings of the US community
  study on the future of particle physics (Snowmass 2021)}},  {\em Eur. Phys.
  J. C} {\bf 83} (2023), no.~1 15, [\href{http://arxiv.org/abs/2203.10811}{{\tt
  arXiv:2203.10811}}].

\bibitem{Berryman:2015nua}
J.~M. Berryman, A.~de~Gouv\^ea, K.~J. Kelly, and A.~Kobach, {\it {Sterile
  neutrino at the Deep Underground Neutrino Experiment}},  {\em Phys. Rev. D}
  {\bf 92} (2015), no.~7 073012, [\href{http://arxiv.org/abs/1507.03986}{{\tt
  arXiv:1507.03986}}].

\bibitem{Agarwalla:2016mrc}
S.~K. Agarwalla, S.~S. Chatterjee, A.~Dasgupta, and A.~Palazzo, {\it {Discovery
  Potential of T2K and NOvA in the Presence of a Light Sterile Neutrino}},
  {\em JHEP} {\bf 02} (2016) 111, [\href{http://arxiv.org/abs/1601.05995}{{\tt
  arXiv:1601.05995}}].

\bibitem{Agarwalla:2016xxa}
S.~K. Agarwalla, S.~S. Chatterjee, and A.~Palazzo, {\it {Physics Reach of DUNE
  with a Light Sterile Neutrino}},  {\em JHEP} {\bf 09} (2016) 016,
  [\href{http://arxiv.org/abs/1603.03759}{{\tt arXiv:1603.03759}}].

\bibitem{Agarwalla:2016xlg}
S.~K. Agarwalla, S.~S. Chatterjee, and A.~Palazzo, {\it {Octant of
  $\theta_{23}$ in danger with a light sterile neutrino}},  {\em Phys. Rev.
  Lett.} {\bf 118} (2017), no.~3 031804,
  [\href{http://arxiv.org/abs/1605.04299}{{\tt arXiv:1605.04299}}].

\bibitem{Agarwalla:2018nlx}
S.~K. Agarwalla, S.~S. Chatterjee, and A.~Palazzo, {\it {Signatures of a Light
  Sterile Neutrino in T2HK}},  {\em JHEP} {\bf 04} (2018) 091,
  [\href{http://arxiv.org/abs/1801.04855}{{\tt arXiv:1801.04855}}].

\bibitem{KumarAgarwalla:2019blx}
S.~Kumar~Agarwalla, S.~S. Chatterjee, and A.~Palazzo, {\it {Physics potential
  of ESS$\nu$SB in the presence of a light sterile neutrino}},  {\em JHEP} {\bf
  12} (2019) 174, [\href{http://arxiv.org/abs/1909.13746}{{\tt
  arXiv:1909.13746}}].

\bibitem{Coloma:2015kiu}
P.~Coloma, {\it {Non-Standard Interactions in propagation at the Deep
  Underground Neutrino Experiment}},  {\em JHEP} {\bf 03} (2016) 016,
  [\href{http://arxiv.org/abs/1511.06357}{{\tt arXiv:1511.06357}}].

\bibitem{Agarwalla:2016fkh}
S.~K. Agarwalla, S.~S. Chatterjee, and A.~Palazzo, {\it {Degeneracy between
  $\theta_{23}$ octant and neutrino non-standard interactions at DUNE}},  {\em
  Phys. Lett.} {\bf B762} (2016) 64--71,
  [\href{http://arxiv.org/abs/1607.01745}{{\tt arXiv:1607.01745}}].

\bibitem{Escrihuela:2016ube}
F.~J. Escrihuela, D.~V. Forero, O.~G. Miranda, M.~T\'ortola, and J.~W.~F.
  Valle, {\it {Probing CP violation with non-unitary mixing in long-baseline
  neutrino oscillation experiments: DUNE as a case study}},  {\em New J. Phys.}
  {\bf 19} (2017), no.~9 093005, [\href{http://arxiv.org/abs/1612.07377}{{\tt
  arXiv:1612.07377}}].

\bibitem{Agarwalla:2021owd}
S.~K. Agarwalla, S.~Das, A.~Giarnetti, and D.~Meloni, {\it {Model-independent
  constraints on non-unitary neutrino mixing from high-precision long-baseline
  experiments}},  {\em JHEP} {\bf 07} (2022) 121,
  [\href{http://arxiv.org/abs/2111.00329}{{\tt arXiv:2111.00329}}].

\bibitem{Chatterjee:2015gta}
S.~S. Chatterjee, A.~Dasgupta, and S.~K. Agarwalla, {\it {Exploring
  Flavor-Dependent Long-Range Forces in Long-Baseline Neutrino Oscillation
  Experiments}},  {\em JHEP} {\bf 12} (2015) 167,
  [\href{http://arxiv.org/abs/1509.03517}{{\tt arXiv:1509.03517}}].

\bibitem{Choubey:2017dyu}
S.~Choubey, S.~Goswami, and D.~Pramanik, {\it {A study of invisible neutrino
  decay at DUNE and its effects on $\theta_{23}$ measurement}},  {\em JHEP}
  {\bf 02} (2018) 055, [\href{http://arxiv.org/abs/1705.05820}{{\tt
  arXiv:1705.05820}}].

\bibitem{Coloma:2017zpg}
P.~Coloma and O.~L.~G. Peres, {\it {Visible neutrino decay at DUNE}},
  \href{http://arxiv.org/abs/1705.03599}{{\tt arXiv:1705.03599}}.

\bibitem{Barenboim:2018ctx}
G.~Barenboim, M.~Masud, C.~A. Ternes, and M.~T\'ortola, {\it {Exploring the
  intrinsic Lorentz-violating parameters at DUNE}},  {\em Phys. Lett. B} {\bf
  788} (2019) 308--315, [\href{http://arxiv.org/abs/1805.11094}{{\tt
  arXiv:1805.11094}}].

\bibitem{KumarAgarwalla:2019gdj}
S.~K. Agarwalla and M.~Masud, {\it {Can Lorentz invariance violation affect the
  sensitivity of deep underground neutrino experiment?}},  {\em Eur. Phys. J.
  C} {\bf 80} (2020), no.~8 716, [\href{http://arxiv.org/abs/1912.13306}{{\tt
  arXiv:1912.13306}}].

\bibitem{Fiza:2022xfw}
N.~Fiza, N.~R. Khan~Chowdhury, and M.~Masud, {\it {Investigating Lorentz
  Violation with the long baseline experiment P2O}},
  \href{http://arxiv.org/abs/2206.14018}{{\tt arXiv:2206.14018}}.

\bibitem{LSND:2005oop}
{\bf LSND} Collaboration, L.~B. Auerbach et~al., {\it {Tests of Lorentz
  violation in $\bar\nu_\mu \to \bar\nu_e$ oscillations}},  {\em Phys. Rev. D}
  {\bf 72} (2005) 076004, [\href{http://arxiv.org/abs/hep-ex/0506067}{{\tt
  hep-ex/0506067}}].

\bibitem{MINOS:2008fnv}
{\bf MINOS} Collaboration, P.~Adamson et~al., {\it {Testing Lorentz Invariance
  and CPT Conservation with NuMI Neutrinos in the MINOS Near Detector}},  {\em
  Phys. Rev. Lett.} {\bf 101} (2008) 151601,
  [\href{http://arxiv.org/abs/0806.4945}{{\tt arXiv:0806.4945}}].

\bibitem{MINOS:2010kat}
{\bf MINOS} Collaboration, P.~Adamson et~al., {\it {A Search for Lorentz
  Invariance and CPT Violation with the MINOS Far Detector}},  {\em Phys. Rev.
  Lett.} {\bf 105} (2010) 151601, [\href{http://arxiv.org/abs/1007.2791}{{\tt
  arXiv:1007.2791}}].

\bibitem{MINOS:2012ozn}
{\bf MINOS} Collaboration, P.~Adamson et~al., {\it {Search for Lorentz
  invariance and CPT violation with muon antineutrinos in the MINOS Near
  Detector}},  {\em Phys. Rev. D} {\bf 85} (2012) 031101,
  [\href{http://arxiv.org/abs/1201.2631}{{\tt arXiv:1201.2631}}].

\bibitem{MiniBooNE:2011pix}
{\bf MiniBooNE} Collaboration, A.~A. Aguilar-Arevalo et~al., {\it {Test of
  Lorentz and CPT violation with Short Baseline Neutrino Oscillation
  Excesses}},  {\em Phys. Lett. B} {\bf 718} (2013) 1303--1308,
  [\href{http://arxiv.org/abs/1109.3480}{{\tt arXiv:1109.3480}}].

\bibitem{DoubleChooz:2012eiq}
{\bf Double Chooz} Collaboration, Y.~Abe et~al., {\it {First Test of Lorentz
  Violation with a Reactor-based Antineutrino Experiment}},  {\em Phys. Rev. D}
  {\bf 86} (2012) 112009, [\href{http://arxiv.org/abs/1209.5810}{{\tt
  arXiv:1209.5810}}].

\bibitem{Super-Kamiokande:2014exs}
{\bf Super-Kamiokande} Collaboration, K.~Abe et~al., {\it {Test of Lorentz
  invariance with atmospheric neutrinos}},  {\em Phys. Rev. D} {\bf 91} (2015),
  no.~5 052003, [\href{http://arxiv.org/abs/1410.4267}{{\tt arXiv:1410.4267}}].

\bibitem{IceCube:2010fyu}
{\bf IceCube} Collaboration, R.~Abbasi et~al., {\it {Search for a
  Lorentz-violating sidereal signal with atmospheric neutrinos in IceCube}},
  {\em Phys. Rev. D} {\bf 82} (2010) 112003,
  [\href{http://arxiv.org/abs/1010.4096}{{\tt arXiv:1010.4096}}].

\bibitem{Abe:2017eot}
{\bf T2K} Collaboration, K.~Abe et~al., {\it {Search for Lorentz and CPT
  violation using sidereal time dependence of neutrino flavor transitions over
  a short baseline}},  {\em Phys. Rev. D} {\bf 95} (2017), no.~11 111101,
  [\href{http://arxiv.org/abs/1703.01361}{{\tt arXiv:1703.01361}}].

\bibitem{Dighe:2008bu}
A.~Dighe and S.~Ray, {\it {CPT violation in long baseline neutrino experiments:
  A Three flavor analysis}},  {\em Phys. Rev. D} {\bf 78} (2008) 036002,
  [\href{http://arxiv.org/abs/0802.0121}{{\tt arXiv:0802.0121}}].

\bibitem{Barenboim:2009ts}
G.~Barenboim and J.~D. Lykken, {\it {MINOS and CPT-violating neutrinos}},  {\em
  Phys. Rev. D} {\bf 80} (2009) 113008,
  [\href{http://arxiv.org/abs/0908.2993}{{\tt arXiv:0908.2993}}].

\bibitem{Rebel:2013vc}
B.~Rebel and S.~Mufson, {\it {The Search for Neutrino-Antineutrino Mixing
  Resulting from Lorentz Invariance Violation using neutrino interactions in
  MINOS}},  {\em Astropart. Phys.} {\bf 48} (2013) 78--81,
  [\href{http://arxiv.org/abs/1301.4684}{{\tt arXiv:1301.4684}}].

\bibitem{Diaz:2015dxa}
J.~S. Diaz, {\it {Correspondence between nonstandard interactions and CPT
  violation in neutrino oscillations}},
  \href{http://arxiv.org/abs/1506.01936}{{\tt arXiv:1506.01936}}.

\bibitem{deGouvea:2017yvn}
A.~de~Gouv\^ea and K.~J. Kelly, {\it {Neutrino vs. Antineutrino Oscillation
  Parameters at DUNE and Hyper-Kamiokande}},  {\em Phys. Rev. D} {\bf 96}
  (2017), no.~9 095018, [\href{http://arxiv.org/abs/1709.06090}{{\tt
  arXiv:1709.06090}}].

\bibitem{Barenboim:2017ewj}
G.~Barenboim, C.~A. Ternes, and M.~T\'ortola, {\it {Neutrinos, DUNE and the
  world best bound on CPT invariance}},  {\em Phys. Lett. B} {\bf 780} (2018)
  631--637, [\href{http://arxiv.org/abs/1712.01714}{{\tt arXiv:1712.01714}}].

\bibitem{Barenboim:2018lpo}
G.~Barenboim, C.~A. Ternes, and M.~T\'ortola, {\it {New physics vs new
  paradigms: distinguishing CPT violation from NSI}},  {\em Eur. Phys. J. C}
  {\bf 79} (2019), no.~5 390, [\href{http://arxiv.org/abs/1804.05842}{{\tt
  arXiv:1804.05842}}].

\bibitem{Majhi:2019tfi}
R.~Majhi, S.~Chembra, and R.~Mohanta, {\it {Exploring the effect of Lorentz
  invariance violation with the currently running long-baseline experiments}},
  {\em Eur. Phys. J. C} {\bf 80} (2020), no.~5 364,
  [\href{http://arxiv.org/abs/1907.09145}{{\tt arXiv:1907.09145}}].

\bibitem{Majhi:2022fed}
R.~Majhi, D.~K. Singha, M.~Ghosh, and R.~Mohanta, {\it {Distinguishing
  Non-Standard Interaction and Lorentz Invariance Violation at Protvino to
  Super-ORCA experiment}},  \href{http://arxiv.org/abs/2212.07244}{{\tt
  arXiv:2212.07244}}.

\bibitem{Giunti:2010zs}
C.~Giunti and M.~Laveder, {\it {Hint of CPT Violation in Short-Baseline
  Electron Neutrino Disappearance}},  {\em Phys. Rev. D} {\bf 82} (2010)
  113009, [\href{http://arxiv.org/abs/1008.4750}{{\tt arXiv:1008.4750}}].

\bibitem{Datta:2003dg}
A.~Datta, R.~Gandhi, P.~Mehta, and S.~U. Sankar, {\it {Atmospheric neutrinos as
  a probe of CPT and Lorentz violation}},  {\em Phys. Lett. B} {\bf 597} (2004)
  356--361, [\href{http://arxiv.org/abs/hep-ph/0312027}{{\tt hep-ph/0312027}}].

\bibitem{Chatterjee:2014oda}
A.~Chatterjee, R.~Gandhi, and J.~Singh, {\it {Probing Lorentz and CPT Violation
  in a Magnetized Iron Detector using Atmospheric Neutrinos}},  {\em JHEP} {\bf
  1406} (2014) 045, [\href{http://arxiv.org/abs/1402.6265}{{\tt
  arXiv:1402.6265}}].

\bibitem{SinghKoranga:2014mxh}
B.~Singh~Koranga and P.~Khurana, {\it {CPT Violation in Atmospheric Neutrino
  Oscillation: A Two Flavour Matter Effects}},  {\em Int. J. Theor. Phys.} {\bf
  53} (2014), no.~11 3737--3743.

\bibitem{Sahoo:2021dit}
S.~Sahoo, A.~Kumar, and S.~K. Agarwalla, {\it {Probing Lorentz Invariance
  Violation with atmospheric neutrinos at INO-ICAL}},  {\em JHEP} {\bf 03}
  (2022) 050, [\href{http://arxiv.org/abs/2110.13207}{{\tt arXiv:2110.13207}}].

\bibitem{Sahoo:2022rns}
S.~Sahoo, A.~Kumar, S.~K. Agarwalla, and A.~Dighe, {\it {Core-passing
  atmospheric neutrinos: a unique probe to discriminate between Lorentz
  violation and non-standard interactions}},
  \href{http://arxiv.org/abs/2205.05134}{{\tt arXiv:2205.05134}}.

\bibitem{Diaz:2016fqd}
J.~S. Diaz and T.~Schwetz, {\it {Limits on CPT violation from solar
  neutrinos}},  {\em Phys. Rev. D} {\bf 93} (2016), no.~9 093004,
  [\href{http://arxiv.org/abs/1603.04468}{{\tt arXiv:1603.04468}}].

\bibitem{Hooper:2005jp}
D.~Hooper, D.~Morgan, and E.~Winstanley, {\it {Lorentz and CPT invariance
  violation in high-energy neutrinos}},  {\em Phys. Rev. D} {\bf 72} (2005)
  065009, [\href{http://arxiv.org/abs/hep-ph/0506091}{{\tt hep-ph/0506091}}].

\bibitem{Tomar:2015fha}
G.~Tomar, S.~Mohanty, and S.~Pakvasa, {\it {Lorentz Invariance Violation and
  IceCube Neutrino Events}},  {\em JHEP} {\bf 11} (2015) 022,
  [\href{http://arxiv.org/abs/1507.03193}{{\tt arXiv:1507.03193}}].

\bibitem{Liao:2017yuy}
J.~Liao and D.~Marfatia, {\it {IceCube\textquoteright{}s astrophysical neutrino
  energy spectrum from CPT violation}},  {\em Phys. Rev. D} {\bf 97} (2018),
  no.~4 041302, [\href{http://arxiv.org/abs/1711.09266}{{\tt
  arXiv:1711.09266}}].

\bibitem{KATRIN:2022qou}
{\bf KATRIN} Collaboration, M.~Aker et~al., {\it {Search for Lorentz-Invariance
  Violation with the first KATRIN data}},
  \href{http://arxiv.org/abs/2207.06326}{{\tt arXiv:2207.06326}}.

\bibitem{Crivellin:2020oov}
A.~Crivellin, F.~Kirk, and M.~Schreck, {\it {Implications of $SU(2)_{L}$ gauge
  invariance for constraints on Lorentz violation}},  {\em JHEP} {\bf 04}
  (2021) 082, [\href{http://arxiv.org/abs/2009.01247}{{\tt arXiv:2009.01247}}].

\bibitem{Kostelecky:2008ts}
V.~A. Kosteleck\'y and N.~Russell, {\it Data tables for lorentz and $cpt$
  violation},  {\em Rev. Mod. Phys.} {\bf 83} (Mar, 2011) 11--31.

\bibitem{Polyakov:1987ez}
A.~M. Polyakov, {\em {Gauge Fields and Strings}}, vol.~3 of {\em {Contemporary
  concepts in physics}}.
\newblock {Harwood Academic Publishers}, 1987.

\bibitem{Kostelecky:1988zi}
V.~A. Kostelecky and S.~Samuel, {\it {Spontaneous Breaking of Lorentz Symmetry
  in String Theory}},  {\em Phys. Rev. D} {\bf 39} (1989) 683.

\bibitem{Kostelecky:1989jp}
V.~A. Kostelecky and S.~Samuel, {\it {Phenomenological Gravitational
  Constraints on Strings and Higher Dimensional Theories}},  {\em Phys. Rev.
  Lett.} {\bf 63} (1989) 224.

\bibitem{Kostelecky:1990pe}
V.~A. Kosteleck\'y and S.~Samuel, {\it Photon and graviton masses in string
  theories},  {\em Phys. Rev. Lett.} {\bf 66} (Apr, 1991) 1811--1814.

\bibitem{Kostelecky:1991ak}
V.~A. Kostelecky and R.~Potting, {\it {CPT and strings}},  {\em Nucl. Phys. B}
  {\bf 359} (1991) 545--570.

\bibitem{Kostelecky:1995qk}
V.~A. Kostelecky and R.~Potting, {\it {Expectation values, Lorentz invariance,
  and CPT in the open bosonic string}},  {\em Phys. Lett. B} {\bf 381} (1996)
  89--96, [\href{http://arxiv.org/abs/hep-th/9605088}{{\tt hep-th/9605088}}].

\bibitem{Kostelecky:1999mu}
V.~A. Kostelecky, M.~Perry, and R.~Potting, {\it {Off-shell structure of the
  string sigma model}},  {\em Phys. Rev. Lett.} {\bf 84} (2000) 4541--4544,
  [\href{http://arxiv.org/abs/hep-th/9912243}{{\tt hep-th/9912243}}].

\bibitem{Kostelecky:2000hz}
V.~A. Kostelecky and R.~Potting, {\it {Analytical construction of a
  nonperturbative vacuum for the open bosonic string}},  {\em Phys. Rev. D}
  {\bf 63} (2001) 046007, [\href{http://arxiv.org/abs/hep-th/0008252}{{\tt
  hep-th/0008252}}].

\bibitem{Gambini:1998it}
R.~Gambini and J.~Pullin, {\it {Nonstandard optics from quantum space-time}},
  {\em Phys. Rev. D} {\bf 59} (1999) 124021,
  [\href{http://arxiv.org/abs/gr-qc/9809038}{{\tt gr-qc/9809038}}].

\bibitem{Alfaro:2002xz}
J.~Alfaro, H.~A. Morales-Tecotl, and L.~F. Urrutia, {\it {Quantum gravity and
  spin 1/2 particles effective dynamics}},  {\em Phys. Rev. D} {\bf 66} (2002)
  124006, [\href{http://arxiv.org/abs/hep-th/0208192}{{\tt hep-th/0208192}}].

\bibitem{Sudarsky:2002ue}
D.~Sudarsky, L.~Urrutia, and H.~Vucetich, {\it {New observational bounds to
  quantum gravity signals}},  {\em Phys. Rev. Lett.} {\bf 89} (2002) 231301,
  [\href{http://arxiv.org/abs/gr-qc/0204027}{{\tt gr-qc/0204027}}].

\bibitem{Amelino-Camelia:2002aqz}
G.~Amelino-Camelia, {\it {Quantum gravity phenomenology: Status and
  prospects}},  {\em Mod. Phys. Lett. A} {\bf 17} (2002) 899--922,
  [\href{http://arxiv.org/abs/gr-qc/0204051}{{\tt gr-qc/0204051}}].

\bibitem{Ng:2003jk}
Y.~J. Ng, {\it {Selected topics in Planck scale physics}},  {\em Mod. Phys.
  Lett. A} {\bf 18} (2003) 1073--1098,
  [\href{http://arxiv.org/abs/gr-qc/0305019}{{\tt gr-qc/0305019}}].

\bibitem{Colladay:1998fq}
D.~Colladay and V.~A. Kostelecky, {\it {Lorentz violating extension of the
  standard model}},  {\em Phys. Rev. D} {\bf 58} (1998) 116002,
  [\href{http://arxiv.org/abs/hep-ph/9809521}{{\tt hep-ph/9809521}}].

\bibitem{Kostelecky:2003fs}
V.~A. Kostelecky, {\it {Gravity, Lorentz violation, and the standard model}},
  {\em Phys. Rev. D} {\bf 69} (2004) 105009,
  [\href{http://arxiv.org/abs/hep-th/0312310}{{\tt hep-th/0312310}}].

\bibitem{Colladay:1996iz}
D.~Colladay and V.~A. Kostelecky, {\it {CPT violation and the standard model}},
   {\em Phys. Rev. D} {\bf 55} (1997) 6760--6774,
  [\href{http://arxiv.org/abs/hep-ph/9703464}{{\tt hep-ph/9703464}}].

\bibitem{Kostelecky:2000mm}
V.~A. Kostelecky and R.~Lehnert, {\it {Stability, causality, and Lorentz and
  CPT violation}},  {\em Phys. Rev. D} {\bf 63} (2001) 065008,
  [\href{http://arxiv.org/abs/hep-th/0012060}{{\tt hep-th/0012060}}].

\bibitem{Kostelecky:2003cr}
V.~A. Kostelecky and M.~Mewes, {\it {Lorentz and CPT violation in neutrinos}},
  {\em Phys. Rev. D} {\bf 69} (2004) 016005,
  [\href{http://arxiv.org/abs/hep-ph/0309025}{{\tt hep-ph/0309025}}].

\bibitem{Bluhm:2005uj}
R.~Bluhm, {\it {Overview of the SME: Implications and phenomenology of Lorentz
  violation}},  {\em Lect. Notes Phys.} {\bf 702} (2006) 191--226,
  [\href{http://arxiv.org/abs/hep-ph/0506054}{{\tt hep-ph/0506054}}].

\bibitem{Kostelecky:2011gq}
A.~Kostelecky and M.~Mewes, {\it {Neutrinos with Lorentz-violating operators of
  arbitrary dimension}},  {\em Phys. Rev. D} {\bf 85} (2012) 096005,
  [\href{http://arxiv.org/abs/1112.6395}{{\tt arXiv:1112.6395}}].

\bibitem{Antonelli:2020nhn}
V.~Antonelli, L.~Miramonti, and M.~D.~C. Torri, {\it {Phenomenological Effects
  of CPT and Lorentz Invariance Violation in Particle and Astroparticle
  Physics}},  {\em Symmetry} {\bf 12} (2020), no.~11 1821,
  [\href{http://arxiv.org/abs/2110.09185}{{\tt arXiv:2110.09185}}].

\bibitem{IceCube:2017qyp}
{\bf IceCube} Collaboration, M.~G. Aartsen et~al., {\it {Neutrino
  Interferometry for High-Precision Tests of Lorentz Symmetry with IceCube}},
  {\em Nature Phys.} {\bf 14} (2018), no.~9 961--966,
  [\href{http://arxiv.org/abs/1709.03434}{{\tt arXiv:1709.03434}}].

\bibitem{Diaz:2009qk}
J.~S. Diaz, V.~A. Kostelecky, and M.~Mewes, {\it {Perturbative Lorentz and CPT
  violation for neutrino and antineutrino oscillations}},  {\em Phys. Rev. D}
  {\bf 80} (2009) 076007, [\href{http://arxiv.org/abs/0908.1401}{{\tt
  arXiv:0908.1401}}].

\bibitem{Kikuchi:2008vq}
T.~Kikuchi, H.~Minakata, and S.~Uchinami, {\it {Perturbation Theory of Neutrino
  Oscillation with Nonstandard Neutrino Interactions}},  {\em JHEP} {\bf 03}
  (2009) 114, [\href{http://arxiv.org/abs/0809.3312}{{\tt arXiv:0809.3312}}].

\bibitem{Esteban:2020cvm}
I.~Esteban, M.~C. Gonzalez-Garcia, M.~Maltoni, T.~Schwetz, and A.~Zhou, {\it
  {The fate of hints: updated global analysis of three-flavor neutrino
  oscillations}},  {\em JHEP} {\bf 09} (2020) 178,
  [\href{http://arxiv.org/abs/2007.14792}{{\tt arXiv:2007.14792}}].

\bibitem{Huber:2004ka}
P.~Huber, M.~Lindner, and W.~Winter, {\it {Simulation of long-baseline neutrino
  oscillation experiments with GLoBES (General Long Baseline Experiment
  Simulator)}},  {\em Comput.Phys.Commun.} {\bf 167} (2005) 195,
  [\href{http://arxiv.org/abs/hep-ph/0407333}{{\tt hep-ph/0407333}}].

\bibitem{Huber:2007ji}
P.~Huber, J.~Kopp, M.~Lindner, M.~Rolinec, and W.~Winter, {\it {New features in
  the simulation of neutrino oscillation experiments with GLoBES 3.0: General
  Long Baseline Experiment Simulator}},  {\em Comput.Phys.Commun.} {\bf 177}
  (2007) 432--438, [\href{http://arxiv.org/abs/hep-ph/0701187}{{\tt
  hep-ph/0701187}}].

\bibitem{Kopp:2007ne}
J.~Kopp, M.~Lindner, T.~Ota, and J.~Sato, {\it {Non-standard neutrino
  interactions in reactor and superbeam experiments}},  {\em Phys. Rev. D} {\bf
  77} (2008) 013007, [\href{http://arxiv.org/abs/0708.0152}{{\tt
  arXiv:0708.0152}}].

\bibitem{Blennow:2013oma}
M.~Blennow, P.~Coloma, P.~Huber, and T.~Schwetz, {\it {Quantifying the
  sensitivity of oscillation experiments to the neutrino mass ordering}},  {\em
  JHEP} {\bf 1403} (2014) 028, [\href{http://arxiv.org/abs/1311.1822}{{\tt
  arXiv:1311.1822}}].

\bibitem{JUNO:2022mxj}
{\bf JUNO} Collaboration, A.~Abusleme et~al., {\it {Sub-percent precision
  measurement of neutrino oscillation parameters with JUNO}},  {\em Chin. Phys.
  C} {\bf 46} (2022), no.~12 123001,
  [\href{http://arxiv.org/abs/2204.13249}{{\tt arXiv:2204.13249}}].

\bibitem{deSalas:2020pgw}
P.~F. de~Salas, D.~V. Forero, S.~Gariazzo, P.~Mart\'\i{}nez-Mirav\'e, O.~Mena,
  C.~A. Ternes, M.~T\'ortola, and J.~W.~F. Valle, {\it {2020 global
  reassessment of the neutrino oscillation picture}},  {\em JHEP} {\bf 02}
  (2021) 071, [\href{http://arxiv.org/abs/2006.11237}{{\tt arXiv:2006.11237}}].

\bibitem{Capozzi:2021fjo}
F.~Capozzi, E.~Di~Valentino, E.~Lisi, A.~Marrone, A.~Melchiorri, and
  A.~Palazzo, {\it {Unfinished fabric of the three neutrino paradigm}},  {\em
  Phys. Rev. D} {\bf 104} (2021), no.~8 083031,
  [\href{http://arxiv.org/abs/2107.00532}{{\tt arXiv:2107.00532}}].

\bibitem{T2K:2023smv}
{\bf T2K} Collaboration, M.~A. Ram\'\i{}rez et~al., {\it {Measurements of
  neutrino oscillation parameters from the T2K experiment using
  $3.6\times10^{21}$ protons on target}},
  \href{http://arxiv.org/abs/2303.03222}{{\tt arXiv:2303.03222}}.

\bibitem{Carceller:2023kdz}
{\bf NOvA} Collaboration, J.~M. Carceller, {\it {3-flavour results with NOvA}},
   {\em PoS} {\bf NOW2022} (2023) 015.

\bibitem{Posiadala-Zezula:2022vzn}
{\bf Super-Kamiokande} Collaboration, M.~Posiadala-Zezula, {\it {Atmospheric
  Neutrino Oscillations with the Super-Kamiokande detector}},  {\em PoS} {\bf
  ICHEP2022} (2022) 573.

\bibitem{Mead:2023spo}
{\bf IceCube} Collaboration, J.~V. Mead, {\it {Atmospheric neutrino
  oscillations in IceCube-DeepCore}},  {\em PoS} {\bf NOW2022} (2023) 007.

\bibitem{KM3NeT:2023ncz}
{\bf KM3NeT} Collaboration, S.~Aiello et~al., {\it {Probing invisible neutrino
  decay with KM3NeT/ORCA}},  {\em JHEP} {\bf 04} (2023) 090,
  [\href{http://arxiv.org/abs/2302.02717}{{\tt arXiv:2302.02717}}].

\bibitem{T2K:2001wmr}
{\bf T2K} Collaboration, Y.~Itow et~al., {\it {The JHF-Kamioka neutrino
  project}},  in {\em {3rd Workshop on Neutrino Oscillations and Their Origin
  (NOON 2001)}}, pp.~239--248, 6, 2001.
\newblock \href{http://arxiv.org/abs/hep-ex/0106019}{{\tt hep-ex/0106019}}.

\bibitem{T2K:2011qtm}
{\bf T2K} Collaboration, K.~Abe et~al., {\it {The T2K Experiment}},  {\em Nucl.
  Instrum. Meth. A} {\bf 659} (2011) 106--135,
  [\href{http://arxiv.org/abs/1106.1238}{{\tt arXiv:1106.1238}}].

\bibitem{T2K:2014xyt}
{\bf T2K} Collaboration, K.~Abe et~al., {\it {Neutrino oscillation physics
  potential of the T2K experiment}},  {\em PTEP} {\bf 2015} (2015), no.~4
  043C01, [\href{http://arxiv.org/abs/1409.7469}{{\tt arXiv:1409.7469}}].

\bibitem{Ayres:2002ws}
D.~Ayres, G.~Drake, M.~Goodman, V.~Guarino, T.~Joffe-Minor, et~al., {\it
  {Letter of Intent to build an Off-axis Detector to study numu to nue
  oscillations with the NuMI Neutrino Beam}},
  \href{http://arxiv.org/abs/hep-ex/0210005}{{\tt hep-ex/0210005}}.

\bibitem{NOvA:2004blv}
{\bf NOvA} Collaboration, D.~S. Ayres et~al., {\it {NOvA: Proposal to Build a
  30 Kiloton Off-Axis Detector to Study $\nu_{\mu} \to \nu_e$ Oscillations in
  the NuMI Beamline}},  \href{http://arxiv.org/abs/hep-ex/0503053}{{\tt
  hep-ex/0503053}}.

\bibitem{NOvA:2007rmc}
{\bf NOvA} Collaboration, D.~S. Ayres et~al., {\it {The NOvA Technical Design
  Report}}, .

\bibitem{Patterson:2012zs}
{\bf NOvA} Collaboration, R.~B. Patterson, {\it {The NOvA Experiment: Status
  and Outlook}},  {\em Nucl. Phys. B Proc. Suppl.} {\bf 235-236} (2013)
  151--157, [\href{http://arxiv.org/abs/1209.0716}{{\tt arXiv:1209.0716}}].

\end{thebibliography}\endgroup
\end{document}